\documentclass[manuscript]{aastex}
\slugcomment{To be published, Astrophysical Journal}
\shortauthors{R. T. Gangadhara}
\shorttitle{CIRCULAR POLARIZATION IN PULSARS} 
\shortauthors{R. T. Gangadhara}
\begin{document}    
\title{CIRCULAR POLARIZATION IN PULSARS DUE TO CURVATURE RADIATION}
\author{R. T. Gangadhara}
\affil{Indian Institute of Astrophysics, Bangalore -- 560 034, India\\
{\tt  ganga@iiap.res.in}}
\begin{abstract} 
  The beamed radio emission from relativistic plasma (particles or
  bunches), constrained to move along the curved trajectories, occurs
  in the direction of velocity. We have generalized the coherent
  curvature radiation model to include the detailed geometry of the
  emission region in pulsar magnetosphere, and deduced the
  polarization state in terms of Stokes parameters.  By considering
  both the uniform and modulated emissions, we have simulated a few
  typical pulse profiles. The antisymmetric type of circular
  polarization survives only when there is modulation or discrete
  distribution in the emitting sources.  Our model predicts a {\it
    correlation between the polarization angle swing and sign reversal
    of circular polarization} as a geometric property of the emission
  process.
\end{abstract}    
    
\section{Introduction} 

Pulsars are highly magnetized with predominantly dipolar field
structure. The rotating magnetic field produces a strong induced
electric field that accelerates charged particles off the surface of
the star into a magnetosphere consisting of predominantly dipolar
magnetic field and corotating relativistic pair plasma.  Pulsar radio
emission models assume that radiation emitted tangentially to the
field lines on which plasma is moving. The polarization state of the
emitted radiation is more or less determined by the structure of
magnetic field at the emission spot.  In the general framework of
models in which the radio power is curvature radiation emitted by
charge bunches constrained to follow field lines, the linear
polarization is intrinsic to the emission mechanism and is,
furthermore, a purely geometric property.  Several pulsar researchers
have shown that the properties such as the polarization angle swing
can be explained within the framework of curvature radiation (e.g.,
Radhakrishnan \& Cooke 1969; Sturrock 1971; Ruderman \& Sutherland
1975; Lyne \& Manchester 1988; Rankin 1990, 1993; Blaskiewicz
et~al. 1991).

The radio emission from particle bunches is highly polarized, and the
radiation received by distant observer will be less polarized due to
the incoherent superposition of emissions from different magnetic
field lines (Gil \& Rudnicki 1985). Gil (1986) has argued for the
connection between pulsar emission beams and polarization modes and
suggested that out of two orthogonal polarization modes one
corresponds to core emission and other to the conal emissions. They
are highly linearly polarized and the observed depolarization is due
to superposition of modes at any instant (Gil 1987). By considering a
charged particle moving along the curved trajectory (circular)
confined to the $xz$-plane, Gil and Snakowski (1990a) have deduced the
polarization state of the emitted radiation, and shown the creation of
antisymmetric circular polarization in curvature radiation. By
introducing a phase, as a propagation effect, the difference between
the components of radiation electric field in the directions parallel
and perpendicular to the plane of particle trajectory, Gil and
Snakowski (1990b) have developed a model to explain the depolarization
and polarization angle deviations in subpulses and micropulses. Gil,
Kijak and Zycki (1993) have modeled the single pulse polarization
characteristics of pulsar radiation, and demonstrated that the
deviations of the single pulse position angle from the average are
caused by both propagation and geometrical effects. Mitra, Gil and
Melikdze (2009) by analysing the strong single pulses with highly
polarized subpulses from a set of pulsars, have given a very
conclusive arguments in favor of the coherent curvature radiation
mechanism as the pulsar radio emission mechanism.

By analyzing the average pulse profiles, Radhakrishnan and Rankin
(1990) have identified two most probable types of circular
polarizations, namely, {\it antisymmetric,} where the circular
polarization changes sense near the core region, and {\it symmetric,}
where the circular polarization remains with same sense. They found
that antisymmetric circular polarization is correlated with the
polarization angle swing, and speculate it to be a geometric property
of the emission mechanism.  Han et~al. (1998), by considering the
published mean profiles, found a correlation between the sense of
circular polarization and polarization angle swing in conal double
profiles, and no significant correlation for core components. Further,
You and Han (2006) have reconfirmed these investigations with a larger
data. However, Cordes et~al. (1978) were the first to point out an
association between the position angle of the linear polarization and
the handedness of the circular polarization.

There are two types of claims for the origin of circular polarization:
intrinsic to the emission mechanism (e.g., Michel 1987; Gil \&
Snakowski 1990a, b; Radhakrishnan \& Rankin 1990; Gangadhara 1997) or
generated by the propagation effects (e.g., Cheng \& Ruderman 1979).
Cheng and Ruderman (1979) have suggested that the expected asymmetry
between the positively and negatively charged components of the
magnetoactive plasma in the far magnetosphere of pulsars will convert
linear polarization to circular polarization. Radhakrishnan and Rankin
(1990) have suggested that the propagation origin of antisymmetric
circular polarization is very unlikely but the symmetric circular
polarization appears to be possible. On the other hand, Kazbegi,
Machabeli and Melikidze (1991, 1992) have argued that the cyclotron
instability, rather than the propagation effect, is responsible for
the circular polarization of pulsars. Lyubarskii and Petrova (1999)
considered that the rotation of the magnetosphere gives rise to wave
mode coupling in the polarization-limiting region, which can result in
circular polarization in linearly polarized normal waves.  Melrose and
Luo (2004) discussed possible circular polarization induced by
intrinsically relativistic effects of pulsar plasma. Melrose (2003)
reviewed the properties of intrinsic circular polarization and
circular polarization due to cyclotron absorption and presented a
plausible explanation of circular polarization in terms of propagation
effects in an inhomogeneous birefringent plasma.  In the
multifrequency simultaneous observations we do find the variations in
the single pulse polarization, which may be attributed to the
propagation effects (Karastergiou et~al. 2001, 2002; Karastergiou,
Johnston \& Kramer 2003).

The correlation between the antisymmetric circular polarization and
the polarization angle swing is a geometric property of the emission
processes (Radhakrishnan and Rankin 1990).  By carefully modeling the
polarization state of the radiation in terms of Stokes parameters, it
is possible to construct the geometry of emission region at
multifrequencies. So far, in the purview of curvature radiation only
the polarization angle has been modeled (Radhakrishnan \& Cooke 1969;
Komesaroff 1970) and attempted to fit it with the average radio
profile data (e.g., Lyne \& Manchester 1988).  Instead of circular
trajectories, it is very important to consider the actual dipolar
magnetic field lines, whose curvature radii vary as a function of
altitude, as the radio emission in pulsars is expected to come from
the range of altitude (e.g., Gangadhara \& Gupta 2001; Gupta \&
Gangadhara 2003; Krzeszowski et al. 2009).  In this paper, we develop
a three-dimensional (3D) model for curvature radiation by relativistic
sources accelerated along the dipolar magnetic field lines. We
consider the actual dipolar magnetic field lines (not the circles) in
a slowly rotating (non-rotating) magnetosphere such that the rotation
effects can be ignored. The relativistic plasma (bunch, i.e., a
point-like huge charge) moving along the dipolar magnetic field lines
emits curvature radiation.  We show that {\it our model reproduces
  polarization angle swing of Radhakrishnan and Cooke (1969), and
  predicts that the correlation of antisymmetric circular polarization
  and polarization angle swing is a geometric property of the emission
  process}. Our model is aimed at re-examining the intrinsic
polarization properties of the vacuum single-particle cuvature
radiation, and planed to consider the propagation effects separately
in the subsequent works.  We derive electric fields of the radiation
field in section~2 and construct the Stokes parameters of the
radiation field in section~3. A few model (simulated) profiles are
presented in section~4 depicting the correlation between the
antisymmetric circular polarization and polarization angle swing in
the different cases of viewing geometry parameters.
 
\section{Electric field of curvature radiation}

\begin{figure} 
\epsscale{1} 
\plotone{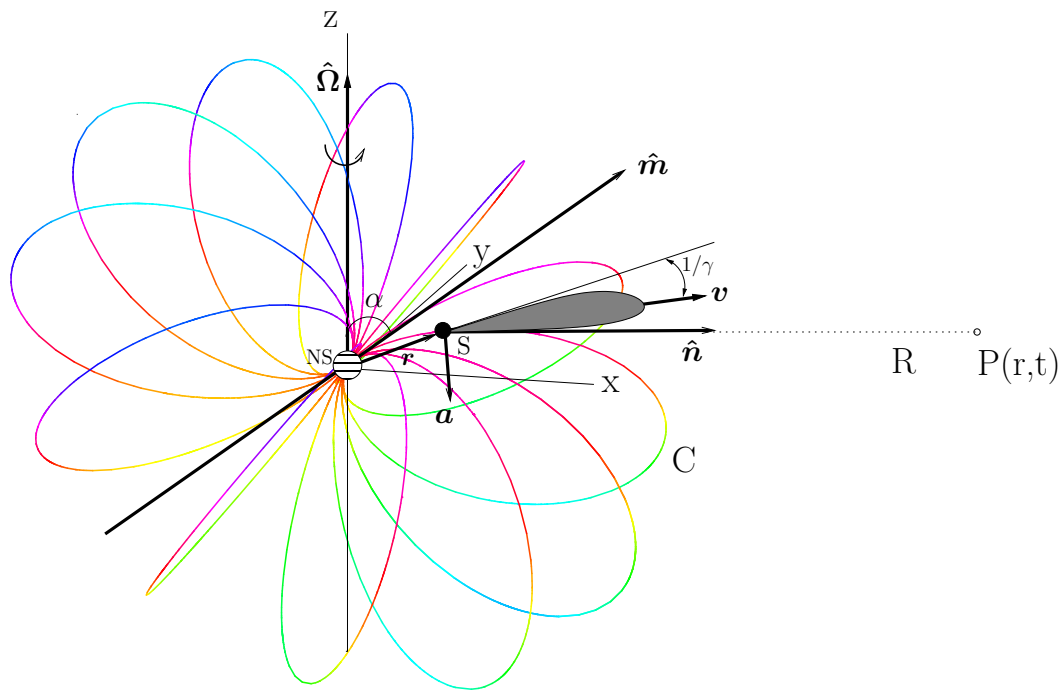}
\caption{Geometry for the calculation of radiation field at P, which
  is at a distance $R$ from the source S. The magnetic axis $\hat m$
  is inclined with respect to rotation axis $\hat \Omega$ by $\alpha.$
  The sight line $\hat n$ impact angle with respect to $\hat m$ is
  $\sigma.$ The colored curves represent the dipolar magnetic field
  lines plotted with $r_e=100$ and azimuthal $(\phi)$ increment of
  $30^\circ$ for each field line, chosen rotation phase $\phi'=0.$ The
  source position vector is \mbox{\boldmath $r$}, velocity is
  \mbox{\boldmath $v$} and acceleration is \mbox{\boldmath $a$}. NS is
  the neutron star and C is an arbitrary field line.}
\label{f1}
\end{figure}

Consider a magnetosphere having dipole magnetic field with an axis
$\hat m$ inclined by an angle $\alpha$ with respect to the rotation
axis $\hat \Omega$ (see Figure~\ref{f1}). We assume that the
magnetosphere is stationary or slowly rotating such that the rotation
effects are negligible. The relativistic pair plasma, generated by
induced electric field followed by pair creation, constrained to move
along the curved dipolar magnetic field lines. The high brightness
temperature of pulsar indicates coherency of the pulsar radiation,
which in tern forces one to postulate the existence of charged
bunches. The formation of bunches in the form of solitons has been
proposed (e.g., Cheng \& Ruderman 1979; Melikidze \& Patarya 1980,
1984) and questioned (e.g., Melrose 1992).  Gil, Lyubarsky and
Melikidze (2004) have generalized the soliton model by including
formation and propagation of the coherent radiation in the
magnetospheric plasma along magnetic field lines. Their results
strongly support coherent curvature radiation by the spark-associated
solitons as a plausible mechanism of pulsar radio emission. Following
these views, we assume that the plasma in the form of bunches moves
along the open field lines of the pulsar magnetosphere.

Consider the source S moving along the magnetic field line C, and
experiencing acceleration ({\boldmath $a$}) in the direction of
curvature vector of the field line. We assume that the source to be a
bunch, which is nothing more than a point-like huge charge.  In
Cartesian coordinates, the position vector of a bunch moving along the
dipolar magnetic field line is given by (see Equation (2) in
Gangadhara 2004, hereafter G04)
\begin{eqnarray} \label{r}
\mbox{\boldmath $r$}& = & r_e \sin^2 \theta\{\cos\theta \cos \phi' \sin
             \alpha+\sin\theta (\cos \alpha \cos \phi \cos \phi'-\sin\phi
             \sin\phi'),\nonumber \\ 
      &    & \cos\phi' \sin\theta \sin \phi+\sin\phi'(\cos\theta \sin \alpha+
             \cos\alpha \cos\phi \sin \theta),\nonumber \\ 
      &    & \cos\alpha \cos\theta-\cos\phi \sin\alpha
   \sin\theta\}~,
\end{eqnarray}
where $r_e$ is the field line constant, and the angles $\theta$ and
$\phi$ are the magnetic colatitude and azimuth, respectively.  Next,
$\phi'$ is the rotation phase and $\alpha$ is the inclination angle of
the magnetic axis. Equation~(\ref{r}) describes the dipolar magnetic
field lines presented in Figure~1.  Then the velocity of bunch is
given by
\begin{equation} 
  \mbox{\boldmath $v$} = \frac{d \mbox{\boldmath $r$}}{dt}= 
  \left(\frac{\partial\mbox{\boldmath $r$}}{\partial\theta}\right)
  \left(\frac{\partial \theta}{\partial t}\right)
  = \left(\frac{\partial \theta}{\partial t}\right)\mbox{\boldmath $b$}~,
\end{equation}
where $\mbox{\boldmath $b$}=\partial\mbox {\boldmath $r$}/\partial
\theta$ is the magnetic field line tangent. Consider the magnetic axis:
\begin{equation}
\hat m = \{\sin\alpha\cos\phi',~\sin\alpha\sin\phi',~\cos\alpha\}.
\end{equation} 

Due to curvature in the field lines, the plasma bunch, a point-like huge
charge, collectively radiates relativistically beamed radiation
in the direction of velocity {\boldmath $v$}. The velocity {\boldmath
  $v$} is parallel to the tangent $\mbox{\boldmath $b$}$ of the
field line.  To receive the beamed emission, the observer's line of sight
($\hat n$) must align with {\boldmath $v$} within the beaming angle
$1/\gamma,$ where $\gamma$ is the Lorentz factor of the bunch. In
other words a distant observer at P receives beamed emission only when
$\hat n\cdot\hat v=\cos\tau\sim 1$ for $\tau\approx 1/\gamma,$ where
$\hat v=\mbox{{\boldmath $v$}}/ \vert\mbox{{\boldmath $v$}}\vert.$ Let
$s$ be the arc length of the field line then $ds=\vert \mbox{\boldmath
  $b$}\vert d\theta, $ where $\vert\mbox{\boldmath
  $b$}\vert=(r_e/\sqrt{2})\sin\theta\sqrt{5+3\cos(2\theta)}, $ and the
magnitude of velocity $v=ds/dt=\kappa\, c,$ where the parameter
$\kappa$ specifies the speed of bunch as a fraction of the speed of
light $c$. Hence, we have
\begin{equation} \label{v}
\mbox{\boldmath $v$}=\kappa\, c\, \hat b~,
\end{equation}
where
\begin{eqnarray} \label{b} \hat b =\mbox{\boldmath
    $b$}/|\mbox{\boldmath $b$}| 
&= &\{\cos\tau \cos \phi' \sin
  \alpha+\sin\tau (\cos \alpha \cos \phi \cos \phi'-\sin\phi
  \sin\phi'),\nonumber \\ 
& & \cos\phi' \sin\tau \sin
  \phi+\sin\phi'(\cos\tau \sin \alpha+ \cos\alpha \cos\phi \sin
  \tau),\nonumber \\ 
& & \cos\alpha \cos\tau-\cos\phi \sin\alpha
  \sin\tau\}~,
\end{eqnarray}
and  $\tau$ is the  angle between $\hat m$ and $\hat b.$ 
In terms of polar angle $\theta,$ the angle $\tau$ given by
\begin{equation}\label{tau}
  \tan\tau  = \frac{\sin\tau}{\cos\tau}
            = \frac{3\sin(2\theta)}{1+3\cos(2\theta)}~,
\end{equation}   
where
\begin{eqnarray} 
\cos\tau & = &\hat b\cdot\hat m=\frac{1+3\cos(2\theta)}
              {\sqrt{10+6\cos(2\theta)}},\nonumber \\
\sin\tau & = &(\hat m \times \hat b)\cdot \hat e_\phi 
           =  \frac{3\sin(2\theta)}{\sqrt{10+6\cos(2\theta)}}
              \nonumber ~,
\end{eqnarray}
and
\begin{equation}
  \hat e_\phi=\{-\cos\alpha\sin\phi\cos\phi'-
              \cos\phi \sin\phi',~\cos\phi\cos\phi'-
  \cos\alpha\sin\phi\sin\phi',~\sin\alpha\sin\phi\}
\end{equation}
is the bi-normal to the field line.  We  solve equation~(\ref{tau})
for $\theta,$ and obtain
\begin{equation}\label{theta}
\cos(2\theta) = \frac{1}{3}(\cos\tau\sqrt{8+\cos^2\tau}-\sin^2\tau)~.
\end{equation}
Hence, from equation~(\ref{v}) it is clear that to receive the
radiation emitted in the direction of tangent $\hat b$ the line sight
line $\hat n$ must line up with it. So, by solving $\hat n\cdot \hat
b=1$ or $\hat n\times\hat b=0,$ we can identify the tangent $\hat b,$
which aligns with $\hat n,$ and hence find the field line curvature
and the coordinates $(\theta, \phi)$ of the emission spot (see
Equations (4), (9), and (11) in G04).  Next, the acceleration of the
bunch is given by
\begin{equation} \label{eq9} 
  \mbox{\boldmath $a$} = \frac{\partial {\mbox{\boldmath $v$}}}{\partial t}
  = \frac{(\kappa\, c)^2}{\vert {\mbox{\boldmath $b$}}\vert}\frac{\partial 
  \hat b}{\partial \theta}= (\kappa\, c)^2 \mbox{\boldmath $k$},
\end{equation}
where $\mbox{\boldmath $k$} = ({1}/{\vert \mbox{\boldmath $b$}\vert})
{\partial \hat b}/{\partial \theta}$ is the curvature (normal) of the
field line. Then the radius of curvature of field line is given by
\begin{equation}
\rho = \frac{1}{\vert \mbox{\boldmath $k$}\vert}
     = \left [2-\frac{8}{3\{3+\cos(2\theta)\}}\right]|\mbox{\boldmath $b$}|~.
\end{equation} 
Therefore, using $\mbox{\boldmath $k$}=\hat k/\rho,$ we can write
\begin{equation} \label{a}
\mbox{\boldmath $a$} = \frac{(\kappa\, c)^2}{\rho} \hat k, 
\end{equation}
where 
\begin{eqnarray} \label{k} 
\hat k & = & \{(\cos\alpha \cos\phi
  \cos\phi'-\sin\phi \sin\phi')
  \cos\tau -\cos \phi '\sin\alpha\sin\tau,\,\, \nonumber \\
  & &(\cos\phi' \sin\phi+\cos\alpha \cos\phi \sin\phi')\cos\tau -
  \sin\alpha \sin\phi'\sin\tau,\,\,\nonumber \\
  & & -\cos\phi \sin\alpha \cos\tau -\cos\alpha\sin\tau\}~.
\end{eqnarray}
The relativistic bunch, i.e., point-like huge charge $q$ that
collectively emits curvature radiation as it accelerates along the
curved trajectory C (see Figure 1). Then the electric field of the
radiation at the observation point P is given by (Jackson 1975):
\begin{equation} \label{Et} 
  \mbox{\boldmath $E$}(\mbox{\boldmath $r$},
  t) = \frac{q}{c}\left[\frac{\hat{n}\times[(\hat{n}- \mbox{\boldmath
        $\beta$})\times{\mbox{\boldmath $\dot\beta$}}]}{R\, \xi^3}
  \right ]_{\rm ret},
\end{equation}
where $\xi=1-\mbox{\boldmath$\beta$}\cdot\hat n,$ $R$ is the distance
from the radiating region to the observer, $\mbox{\boldmath
  $\beta$}=\mbox{\boldmath $v$}/c$ is the velocity, and $\mbox{\boldmath
  $\dot\beta$}=\mbox{\boldmath $a$}/c$ is the acceleration of the bunch.

The radiation emitted by a relativistic bunch has a broad spectrum,
and it can be estimated by taking Fourier transformation of the
electric field of radiation:
\begin{equation}
  \mbox{\boldmath $E$}(\mbox{\boldmath $r$},\omega) = 
           \frac{1}{\sqrt{2\pi}}\int\limits_{-\infty}^{+\infty}
  \mbox{\boldmath $E$}(\mbox{\boldmath $r$},t) e^{i\,\omega t} dt. 
\end{equation}
In equation~(\ref{Et}), ret means evaluated at the retarded time
$t'+R(t')/c = t.$ By changing the variable of integration from $t$
to $t',$ we obtain
\begin{equation}\label{Eo1}
  \mbox{\boldmath $E$}((\mbox{\boldmath $r$},\omega) =
  \frac{1}{\sqrt{2\pi}}\frac{q}{c}\!\!\int
  \limits_{-\infty}^{+\infty}\frac{\hat{n}\times[(\hat{n}-\mbox{\boldmath
        $\beta$}) \times\mbox{\boldmath $\dot\beta$}]}{R\, \xi^2}
    e^{i\omega\{t'+R(t')/c\}} dt',
\end{equation}
where we have used $dt = \xi\, dt'.$ When the observation point is far
away from the region of space where the acceleration occurs, the
propagation vector or the sight line $\hat n$ can be taken to be
constant in time. Furthermore, the distance $R(t')$ can be
approximated as $R(t')\approx R_{\rm 0}-\hat{n}\cdot\mbox{\boldmath
  $r$}(t'),$ where $R_{\rm 0}$ is the distance between the origin O
and the observation point P, and $\mbox{\boldmath $r$}(t')$ is the
position of the bunch relative to O.

\begin{figure}
\epsscale{0.75}
\plotone{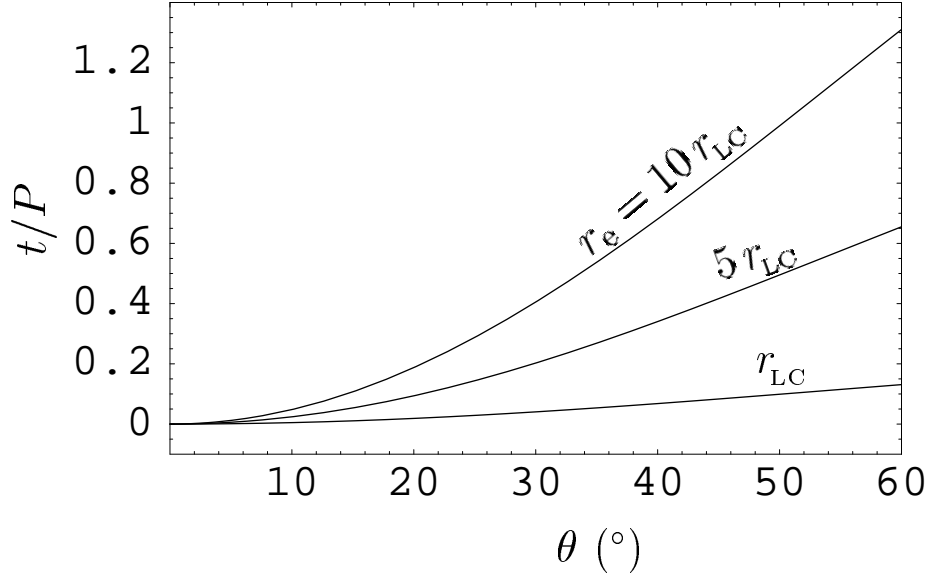}
\caption{Time $t$ is plotted as a function of magnetic
    colatitude $\theta$ of the bunch for different values of field line
    constant $r_{\rm e}.$ The normalization parameter $P$ is the
    pulsar period. Given $\kappa=1.$}
\label{f2}
\end{figure}

Since bunches move with velocity $\kappa\, c$ along the dipolar field
lines, over the incremental time $dt$ the distance (arc length)
covered is $ds = \kappa\, c\, dt = \vert \mbox{\boldmath $b$} \vert
d\theta.$ Therefore, we have
\begin{equation}
 t = \frac{1}{\kappa\, c}\int\vert \mbox{\boldmath $b$} \vert d\theta
   = \frac{r_{\rm e}}{\sqrt{2}\kappa\, c}\int\sin\theta
      \sqrt{5+3 \cos(2 \theta)}\, d\theta ~.
\end{equation}
By choosing $t=0$ at $\theta=0,$ we obtain
\begin{eqnarray}\label{t}
t & = &\frac{r_{\rm e}}{12  \kappa c} \bigg[12+\sqrt{3} \log 
       \left(14+8\sqrt{3}\right)-3 \sqrt{10+6 \cos (2 \theta )} 
       \cos (\theta )- \nonumber \\
  &   & 2 \sqrt{3} \log \left(\sqrt{6} \cos (\theta )+
        \sqrt{5+3 \cos (2 \theta )}\right) \bigg ]~.
\end{eqnarray}
By assuming $\kappa\sim 1$, in Figure~\ref{f2}, we plotted $t$ as a
function of $\theta$ for different $r_{\rm e}.$ It shows time $t$
increases much faster at larger $r_{\rm e}$ than at lower. This is due to
the fact that for a given range of $\theta$ the arc length of the
field line becomes larger at higher $r_{\rm e}.$

Then equation~(\ref{Eo1}) becomes
\begin{equation}\label{Eo2}
\mbox{\boldmath $E$}(\mbox{\boldmath $r$},\omega)\approx \frac{q\,
  {\rm e}^{i\omega R_{\rm 0}/ c}} {\sqrt{2\pi} R_{\rm 0}\kappa c^2}
\!\int\limits_{-\infty}^{+\infty} \vert\mbox{\boldmath $b$}\vert
\frac{\hat{n}\times[(\hat{n}- \mbox{\boldmath $\beta$}) \times
    \mbox{\boldmath $\dot\beta$}]}{\xi^2} {\rm e}^{i\omega \{t-\hat
  n.{\bf r}/c\}} d\theta,
\end{equation}
where the expression for $t$ is given by equation~(\ref{t}). Note that
the prime on the time variable $t$ has been omitted for brevity. The
integration limits have been extended to $\pm\infty$ for mathematical
convenience, as the integrand vanishes for $|\theta-\theta_0|
>1/\gamma.$ At any rotation phase $\phi',$ there exists a magnetic
colatitude $\theta_0$ and a magnetic azimuth $\phi_0$ at which the
field line tangent $\hat b$ exactly align with $\hat n,$ {\it i.e.,}
$\hat b_0\cdot\hat n=1$ and $\tau=\Gamma,$ where $\Gamma$ is the
half-opening angle of the pulsar emission beam centered on $\hat m.$
The expressions for $\theta_0$ and $\phi_0$ are given in G04 (see
Equations (9) and (11)).

The polarization state of the emitted radiation can be determined
using $\mbox{\boldmath $E$}(\omega)$ with the known {\boldmath
  $r$}$(t),$ {\boldmath $\beta$} and {\boldmath $\dot\beta$}. Since
the integral in equation~(\ref{Eo2}) has to be computed over the path
of particle the line of sight $\hat n$ can be chosen without loss of
generality, to lie in the $xz$--plane:
\begin{equation}\label{los}
      \hat n = (\sin\zeta,~0, ~\cos\zeta)~, 
\end{equation}
where $\zeta = \alpha+\sigma$ is the angle between $\hat n$ and $\hat
\Omega,$ and $\sigma$ is the closest impact angle of $\hat n$ with
respect to $\hat m.$

Let 
\begin{equation}
  \mbox{\boldmath $A$} = \frac{1}{\kappa c}\vert\mbox{\boldmath
    $b$}\vert\frac{\hat{n}\times[(\hat{n}- \mbox{\boldmath
        $\beta$})\times \mbox{\boldmath $\dot\beta$}]}{\xi^2}~.
\end{equation}
By substituting for acceleration {\boldmath $\dot\beta=a/$c} from 
equation~(\ref{eq9}), we can reduce it to
\begin{equation}
  \mbox{\boldmath $A$} = \{A_{\rm x},\, A_{\rm y},\, A_{\rm z}\}
  = \frac{\hat{n}\times[(\hat{n}-
    \mbox{\boldmath $\beta$})\times \mbox{\boldmath $N$}]}{\xi^2}~,
\end{equation}
where \mbox{\boldmath $N$} $=\kappa\, \partial \hat b/\partial\theta$=
$\partial\mbox{\boldmath $\beta$}/\partial\theta .$
Using the expression {\boldmath$\beta=v$}$/c$ from equation~(\ref{v})
 and series expanding \mbox{\boldmath $A$} in power
of $\theta$ about $\theta_0$ we obtain
\begin{eqnarray} \label{coe}
A_{\rm x} & = & A_{\rm x0}+ A_{\rm x1} (\theta-\theta_0)+A_{\rm x2} 
            (\theta-\theta_0)^2+A_{\rm x3} 
                (\theta-\theta_0)^3+O[(\theta-\theta_0)^4]~,\nonumber \\
A_{\rm y} & = & A_{\rm y0}+ A_{\rm y1} (\theta-\theta_0)+A_{\rm y2} 
               (\theta-\theta_0)^2+A_{\rm y3} 
                (\theta-\theta_0)^3+O[(\theta-\theta_0)^4]~,\nonumber \\
A_{\rm z} & = & A_{\rm z0}+ A_{\rm z1} (\theta-\theta_0)+A_{\rm z2} 
           (\theta-\theta_0)^2+A_{\rm z3} 
                (\theta-\theta_0)^3+O[(\theta-\theta_0)^4]~,
\end{eqnarray}
where $A_{{\rm x}\it{i}},\, A_{{\rm y}\it{i}}$ and $A_{{\rm z}\it{i}} $
with $i = 0,\,1,\,2,\,3$ are the series expansion coefficients, and their
expressions are given in Appendix~A.

The scalar product between $\hat n$ and {\boldmath $r$} is given by
\begin{eqnarray}
  \hat n\cdot\mbox{\boldmath $r$} 
   & = & r_e\sin^2\theta\Big[\cos\alpha \left(\cos\theta\cos\zeta+
    \cos\phi\cos\phi'\sin\theta\sin\zeta\right)
  -\cos\zeta\cos\phi\sin\alpha\sin\theta + \nonumber \\
  &   & \sin\zeta\left(\cos\theta\cos\phi'\sin\alpha-
    \sin\theta\sin\phi\sin\phi'\right)\Big]~.
\end{eqnarray}  
Next, substituting the expressions of $t$ and $\hat n\cdot
\mbox{\boldmath $r$}$ into the argument of exponential in
equation~(\ref{Eo2}), and series expanding in powers of $\theta$ about
$\theta_0$ we obtain
\begin{equation}\label{exp}
  \omega\left(t-\frac{\hat n\cdot\mbox{\boldmath $r$}}{c}\right)=
  c_{\rm 0}+ c_{\rm 1} (\theta-\theta_0)+c_{\rm 2}
  (\theta-\theta_0)^2+c_{\rm 3}
  (\theta-\theta_0)^3+O[(\theta-\theta_0)^4]~,
\end{equation}
where $c_0$, $c_1,$ $c_2,$ and $c_3$  are the series expansion
coefficients, and their expressions are given in Appendix~A.

Now, by substituting the expressions of equations~(\ref{coe}) and
(\ref{exp}) into equation~(\ref{Eo2}), we obtain the components of
$\mbox{\boldmath $E$}(\omega)=\{E_x(\omega),\, E_y(\omega),\,
E_z(\omega)\}:$
\begin{eqnarray}\label{Ecomp}
E_{\rm x}(\omega)& = & E_0 \!\int\limits_{-\infty}^{+\infty} 
     (A_{\rm x0}+ A_{\rm x1} \,\mu+A_{\rm x2} \,\mu^2+A_{\rm x3} 
\mu^3) 
     { e}^{i(c_{\rm 1} \,\mu+c_{\rm 2} \,\mu^2+c_{\rm 3} \,\mu^3)}d\mu~, 
     \nonumber \\
E_{\rm y}(\omega)& = & E_0  \!\int\limits_{-\infty}^{+\infty} 
    (A_{\rm y0}+ A_{\rm y1} \,\mu+A_{\rm y2} \,\mu^2+A_{\rm y3} \,\mu^3) 
    {e}^{i(c_{\rm 1} \,\mu+c_{\rm 2} \,\mu^2+c_{\rm 3} \,\mu^3)}d\mu ~,
    \nonumber \\
E_{\rm z}(\omega)& = & E_0  \!\int\limits_{-\infty}^{+\infty} 
    (A_{\rm z0}+ A_{\rm z1} \,\mu+A_{\rm z2} \,\mu^2+A_{\rm z3} \,\mu^3)
    {e}^{i(c_{\rm 1} \,\mu+c_{\rm 2} \,\mu^2+c_{\rm 3
} \,\mu^3)}d\mu~,
\end{eqnarray}
where $\mu=\theta-\theta_0$ ~~and 
$$E_0 = \frac{q}{\sqrt{2\pi} R_{\rm 0} c}{ e}^{i[(\omega
  R_{\rm 0}/c)+c_0]}.$$

Now by substituting the integral solutions $S_0,$ $S_1,$ $S_2$ and
$S_3,$ given in Appendix~B, into equation~(\ref{Ecomp}) we obtain
\begin{eqnarray}\label{Ecomp1}
E_{\rm x}(\omega) & = & E_0 (A_{\rm x0}S_0+ A_{\rm x1}S_1+A_{\rm x2}S_2+
                        A_{\rm x3}S_3)~, \nonumber \\
E_{\rm y}(\omega) & = & E_0 (A_{\rm y0}S_0+ A_{\rm y1}S_1+A_{\rm y2}S_2+
                        A_{\rm y3}S_3)~, \nonumber \\
E_{\rm z}(\omega) & = & E_0 (A_{\rm z0}S_0+ A_{\rm z1}S_1+A_{\rm z2}S_2+
                        A_{\rm z3}S_3)~.
\end{eqnarray}

To find the polarization angle of radiation field {\boldmath $E$}, we
need to specify two reference directions perpendicular to the sight
line $\hat n.$ One could be the projected spin axis on the plane of
the sky: $\hat \epsilon_\parallel=(-\cos\zeta,\,0, \,\sin\zeta), $
and then the other direction is specified by
$\hat\epsilon_\perp=\hat\epsilon_\parallel\times\hat n=\hat y,$ where
$\hat y$ is a unit vector parallel to the $y$-axis. Then the components
of {\boldmath $E$} in the directions $\hat \epsilon_\parallel$ and
$\hat \epsilon_\perp$ are given by
\begin{eqnarray}
E_\parallel & = &\hat \epsilon_\parallel\cdot\mbox{\boldmath $E$}
             = -\cos\zeta\, E_{\rm x}+\sin\zeta \,E_{\rm z}~, \nonumber \\
E_\perp    & = &\hat\epsilon_\perp\cdot\mbox{\boldmath $E$} =E_{\rm y}~.
\end{eqnarray}

\begin{figure}
\epsscale{1.  } 
\plotone{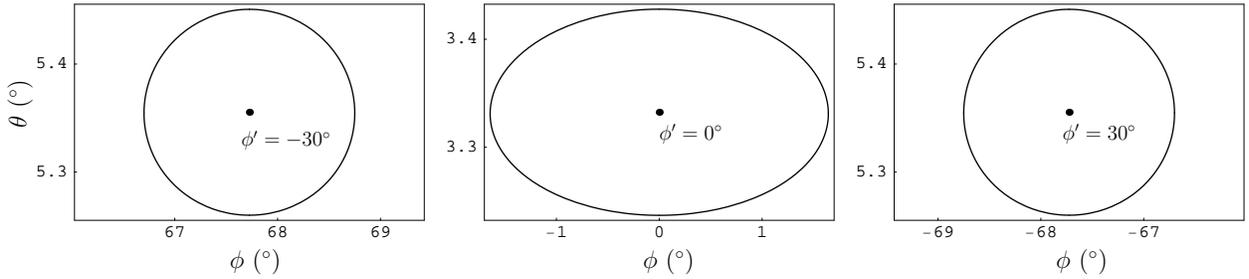}
\caption{Beaming regions specifying the range of magnetic
    colatitude $\theta$ and azimuth $\phi$ at the three selected
    phases $\phi'=-30^\circ,$ $0^\circ$ and $30^\circ.$ The center of
    each region gives the values of $\phi_0$ and $\theta_0.$ Given
    $\alpha=10^\circ,$ $\beta=5^\circ$ and $\gamma=400.$ }
\label{f3}
\end{figure}

\begin{figure}
\epsscale{0.75}
\plotone{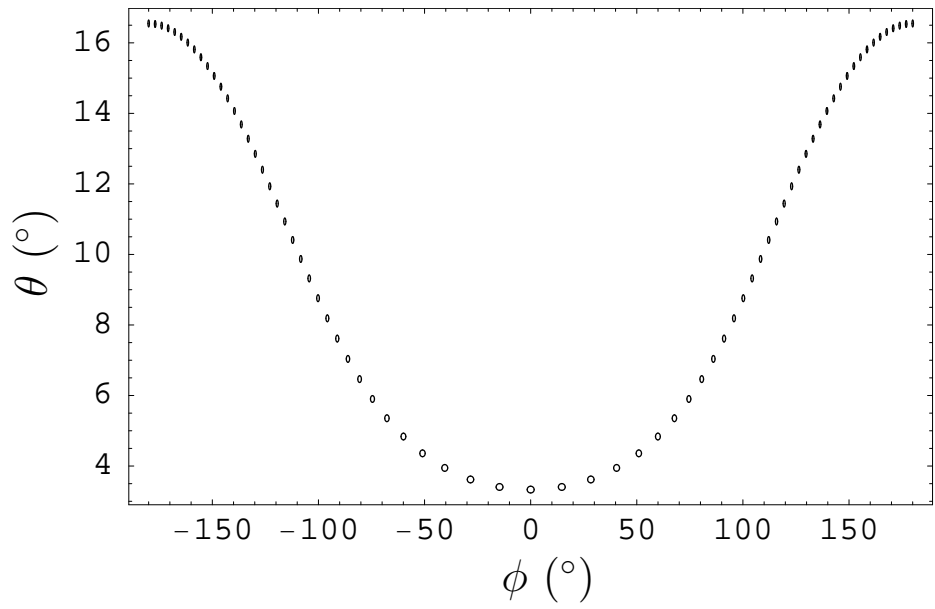}
\caption{Beaming regions specifying the range of magnetic
    colatitude $\theta$ and azimuth $\phi.$ They are plotted for the
    full range of phase: $-180^\circ\leq\phi'\leq 180^\circ$ with a step of
    $5^\circ$. The center of each region gives the values of $\phi_0$
    and $\theta_0.$  Given
    $\alpha=10^\circ,$ $\beta=5^\circ$ and $\gamma=400.$}
\label{f4}
\end{figure}

At any rotation phase $\phi',$ the observer receives the radiation from
all those field lines whose tangents lie within the angle $1/\gamma$
with respect to the sight line $\hat n.$ Let $\eta$ be the angle
between the $\hat b$ and $\hat n,$ then $\cos\eta=\hat b\cdot \hat n,$
and the maximum value of $\eta$ is $1/\gamma.$ Therefore, at
$\phi=\phi_0$ we solve $\cos (1/\gamma)=\hat b\cdot \hat n$ for
$\tau,$ and find the allowed range $(\Gamma-1/\gamma)\leq \tau \leq
(\Gamma+1/\gamma)$ of $\tau$ or $-1/\gamma\leq \eta \leq 1/\gamma$ of
$\eta,$ which in turn allows to one find the range of $\theta$ with the
help of equation~(\ref{theta}). Next, for any given $\eta$ within its
range, we find $\phi$ by solving $\cos\eta=\hat b\cdot \hat n.$ It
gives $(\phi_0-\delta\phi)\leq\phi \leq(\phi_0+\delta\phi),$ where
\begin{equation}\label{dphi}
\cos(\delta\phi)=\frac{ \sin\Gamma[\cos(1/\gamma) \csc(\Gamma+\eta)-
    \cos\Gamma \cot(\Gamma +\eta)]}{(\cos\zeta\sin\alpha-\cos\alpha
    \cos\phi'\sin\zeta)^2+\sin^2\zeta\sin^2\phi'}~.
\end{equation}
Hence by knowing the ranges of $\theta$ and $\phi$ at any given
$\phi',$ we can estimate the contributions to {\boldmath $E$} from all
those field lines, whose tangents lie within the angle $1/\gamma$ with
respect to $\hat n.$ In Figure~\ref{f3}, we have plotted those regions
at three phases; $\phi'=-30^\circ,~0^\circ$ and $30^\circ$ using
$\alpha=10^\circ,$ $\beta=5^\circ,$ and $\gamma=400.$ Note that at the
center of each region $\hat b$ exactly aligns with the sight line,
i.e., $\hat b \cdot \hat n=1.$ Further, in Figure~\ref{f4}, we have
plotted them for $-180^\circ\leq \phi'\leq 180^\circ$ with a step of
$5^\circ$ between the successive regions. We observe that the range of
$\theta$ stays nearly constant (or decreasing negligibly) whereas that
of $\phi$ gets narrower with respect to the increasing $|\phi'|.$
 
\section{Polarization of radiation field} 

To understand the pulsar radio emission, we must model all the Stokes
parameters ($I,$ $Q,$ $U$ and $V$ ---a set of parameters used to
specify the phase and polarization of radiation), and compare with 
observations.  They have been found to offer a very convenient method
for establishing the association between the polarization state of
observed radiation and the geometry of the emitting region.  They are
defined as follows:
\begin{equation}
I  =  E_\parallel E_\parallel^*+E_\perp E_\perp^*~,  \quad
Q  =  E_\parallel E_\parallel^*-E_\perp  E_\perp^*~, \quad
U  =  2\, {\rm Re}[E_\parallel^* E_\perp]~, \quad
V  =  2\, {\rm Im}[E_\parallel^* E_\perp]~.
\end{equation}
The parameter $I$ defines the total intensity, $Q$ and $U$ jointly
define the linear polarization and it's position angle, and $V$ 
describes the circular polarization.

\subsection{Addition of Stokes parameters}\label{stokes_add}
   Let $W_I$ be the energy radiated coherently per unit solid angle
per unit frequency interval per particle bunch (Jackson 1975), then
\begin{eqnarray}\label{eq_E_Ster_nu}
   \frac{d^2 W_I}{d\omega\,d\Omega}= \frac{c\, R_0^2}{2\pi} 
       \!|\mbox{\boldmath $E$}(\omega)|^2\!.
\end{eqnarray}
Since the Stokes parameter $I=E_\parallel {E_\parallel}^*+E_\perp
{E_\perp}^*=\mbox{\boldmath $E$} \cdot \mbox{\boldmath
  $E$}^*=|\mbox{\boldmath $E$}|^2,$ we can rewrite
equation~(\ref{eq_E_Ster_nu}) as
\begin{eqnarray}\label{eq_E_Ster_nu_1} 
  I= |\mbox{\boldmath $E$}|^2 =  \frac{2\pi} {c\, R_0^2} 
      \frac{d^2 W_I}{d\omega\,d\Omega}.
\end{eqnarray}
Similarly, we can express  $Q,$ $U$ and $V$ as
 \begin{eqnarray}\label{eq_stokesa}
 Q &=& \frac{2\pi} {c\, R_0^2} \frac{d^2 W_Q}{d\omega\,d\Omega},\nonumber \\
 U &=& \frac{2\pi} {c\, R_0^2} \frac{d^2 W_U}{d\omega\,d\Omega},\nonumber \\
 V &=& \frac{2\pi} {c\, R_0^2} \frac{d^2 W_V}{d\omega\,d\Omega}~. 
\end{eqnarray}
The net emission, which the observer receives along $\hat n$, will
have contributions from the neighboring field lines, whose tangents
are within the angle $1/\gamma$ with respect to $\hat n.$ Hence the
radiation received at any given phase is the net contribution from a
small tube of field lines having an angular width of about $2/\gamma.$
Thus the radiation in the direction of $\hat n$ should be integrated
over a solid angle $d\Omega = \sin\theta\,d\theta\, d\phi.$ We choose
limits on the angles $\phi$ and $\theta$ such that the integration
over them will cover the solid angular region (beaming region) of
radial width $1/\gamma$ around $\hat n.$ Since $\theta$ and $\phi$ are
orthogonal, choosing them as the variables of integration is
justified.  We assume (i) the width of bunch $\eta_0$ is much smaller
than the wavelength $\lambda$ of the radio waves, so that the
radiation emitted by a bunch is coherent, and (ii) the bunches, within
the beaming region, are closely spaced, so that the net emission
becomes smooth and continuous.

Consider a bunch having $\gamma \sim 400$ emitting radio waves at
frequency $\nu = 600$~MHz at an altitude of about 400~km.  Note that
these values are closer to those estimated in G04 in the case of
PSR~B0329+54.  Then the angular width of the beaming region
corresponding to $2/\gamma$ is $\sim 0.3^\circ,$ which corresponds to
width of $\sim 2$~km at an altitude of 400 km.  For coherence to be
effective the bunch width $w_0 < \lambda.$ Therefore, we choose $w_0 <
50$~cm for $\lambda \sim 50$~cm.  Since these values of $w_0$ are much
smaller than the width of the beaming region ($\sim2$~Km), the Stokes
parameters can be integrated as continuous functions of $\theta$ and
$\phi.$

Let $I_{\rm s}$ be the resultant Stokes intensity parameter then
\begin{eqnarray}\label{Is}
I_{\rm s}&=& \int\,I\,d\Omega ~\nonumber \\
 & =& \int^{\theta_{\rm 0}+\delta\theta}_{\theta_{\rm 0}-\delta\theta}
           \int^{\phi_0+\delta\phi}_{\phi_0-\delta\phi} I\, 
             \sin\theta\,d\theta\,d\phi ~,
\end{eqnarray} 
where $\theta_{\rm 0}$ and $\phi_0$ are the magnetic colatitude
and azimuth of the sight line $\hat n.$
Similarly, for other Stokes parameters, we have
\begin{eqnarray}\label{QUVs}
Q_{\rm s} &=& \int^{\theta_{\rm 0}+\delta\theta}_{\theta_{\rm 0}-\delta\theta}
           \int^{\phi_0+\delta\phi}_{\phi_0-\delta\phi} Q\,
             \sin\theta\,d\theta\,d\phi ~,\nonumber \\
U_{\rm s} &=& \int^{\theta_{\rm 0}+\delta\theta}_{\theta_{\rm 0}-\delta\theta}
           \int^{\phi_0+\delta\phi}_{\phi_0-\delta\phi} U\, 
             \sin\theta\,d\theta\,d\phi ~,\nonumber \\
V_{\rm s} &=& \int^{\theta_{\rm 0}+\delta\theta}_{\theta_{\rm 0}-\delta\theta}
           \int^{\phi_0+\delta\phi}_{\phi_0-\delta\phi} V\, 
             \sin\theta\,d\theta\,d\phi ~.
\end{eqnarray} 
Then the linear polarization is given by  
\begin{equation} 
L_{\rm s}=\sqrt{Q_{\rm s}^2+U_{\rm s}^2},
\end{equation} 
and the corresponding polarization angle is 
\begin{equation}
\psi_{\rm s}=\frac{1}{2} \tan^{-1}\left(\frac{U_{\rm s}}{Q_{\rm s}}\right).
\end{equation} 
\section{Simulation of pulse profiles}

The emission in spin-powered pulsars is mostly of non-thermal origin.
If the radiation field {\boldmath $E$} from different sources does not
bear any phase relation then they are expected to be incoherently
superposed on the observation point.  On the other hand, if there is a
phase relation then they are coherently superposed. From the
observational point of view both the cases are important.

\begin{figure}
\epsscale{1.0}
\plotone{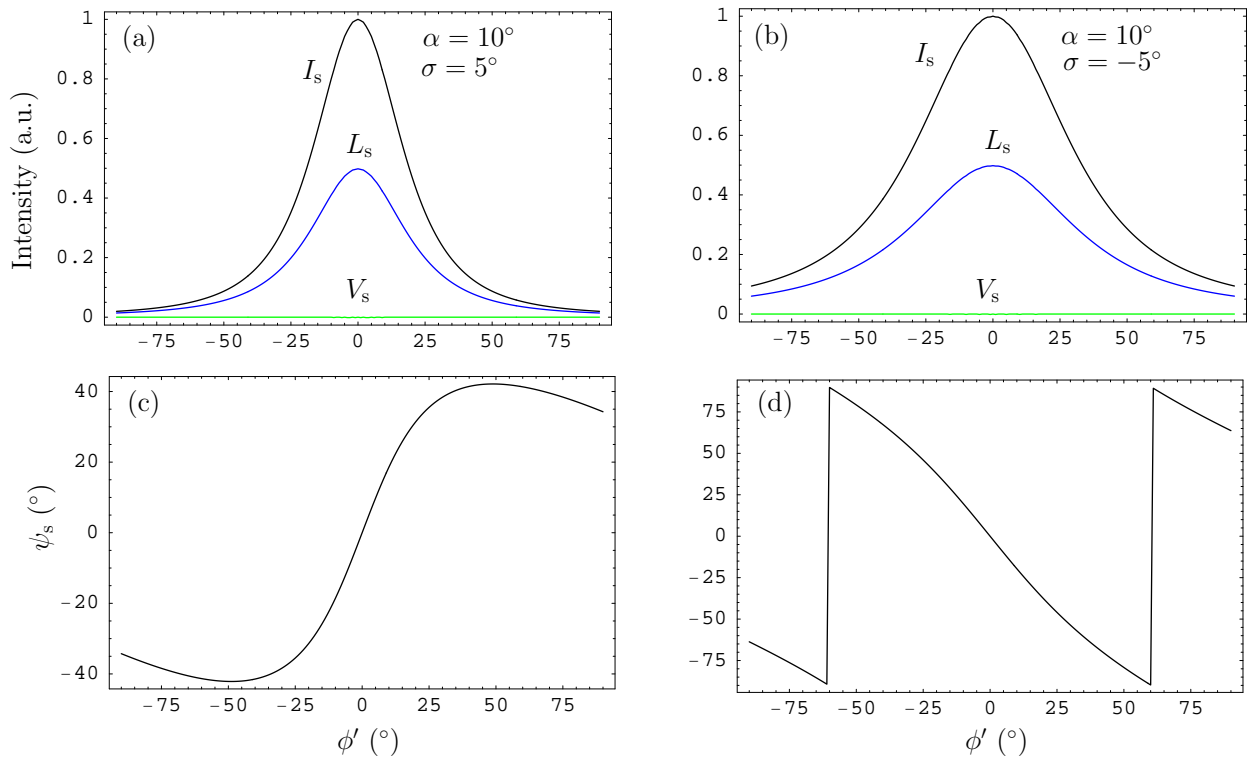}
\caption{Simulated pulse profiles: in panels (a) and (b)
    intensity $(I_{\rm s})$, linear polarization $(L_{\rm s})$ and
    circular polarization $(V_{\rm s})$, and in lower panels (c) and
    (d) the corresponding polarization angle $(\psi_{\rm s})$ curves
    are plotted.  Given $P=1$~s and $\gamma=400.$ Note that profiles
    are normalized with the peak intensity.}
\label{f5}
\end{figure}

By considering the relativistic pair plasma with $\gamma=400$
accelerated along the dipolar field lines of a pulsar with period
$P=1$~s, we computed the polarization parameters and plotted them in
Figure~\ref{f5}. It shows a stronger emission near the meridional
plane, where the beaming region is broader (see Figure~\ref{f3}) and
the radius of curvature $\rho$ goes to a minimum.  The profile of
linear polarization $L_{\rm s}$ resembles the intensity profile,
except for its lower magnitude due to the incoherent addition. To
describe the behaviors of circular polarization $V_{\rm s}$ and
polarization angle $\psi_{\rm s}$, we define the symbols: `$-/+$' for
transition of the right hand circular to left hand circular, `$+/-$'
for left hand circular to right hand circular, `cw' for clockwise
rotation of the polarization angle, and `ccw' for counter clockwise
rotation.

Since the circular polarization $V_{\rm s}$ changes sign as $-/+$ or
$+/-$ as the sight line cuts across the field line, the net circular
polarization goes to zero in an uniform emission due to the addition
with opposite signs.  The polarization angle swings reproduced in
Figure~\ref{f5} are consistent with the rotating vector model of
Radhakrishnan and Cooke (1969). In the case of positive sight line
impact parameter $(\sigma=5^\circ),$ the polarization angle swing is
ccw as the slope $d\psi_{\rm s}/d\phi'>0$ while in the negative case
$(\sigma = -5^\circ),$ it is cw as $d\psi_{\rm s}/d\phi'<0.$

\subsection{Modulation of radio emission}
Pulsar radio emission is believed to come from mostly open magnetic
field lines, whose foot points define the polar cap. The shape of
pulsar profiles indicates that the entire polar cap does not radiate,
only some selected regions radiate, which may be organized into a
central core emission and coaxial conal emissions, which has an
overwhelming support from observations (e.g., Rankin 1990,
1993). Hence the radiating region above the polar is believed to have
a central column of emission (core) and a few coaxial conal regions of
emission (cones) (e.g., Gil \& Krawczyk 1997; Gangadhara \& Gupta
2001; Gupta \& Gangadhara 2003; Dyks, Rudak \& Harding 2004).

\subsubsection{Modulating function}
It is well known that the components of a pulsar profile can be
decomposed into individual Gaussians by fitting one with each of the
subpulse component. For example the components in the pulse profile of
PSR~1706-16 and PSR~2351+61 are fitted with appropriate Gaussians by
Kramer et~al. (1994).  When the line--of--sight crosses the emission
region, it encounters a pattern in intensity due to Gaussian
modulation in the azimuthal direction. Because of the Gaussian
modulation in the azimuthal direction, the intensity becomes
nonuniform in the polar directions too.  These arguments indicate that
a Gaussian-like intensity modulation exists in the polar directions
too. So, we assume that the emission region of a pulse component has
an intensity modulation both in the azimuthal directions.  Hence we
define a modulation function $f$ for a pulse component as
\begin{equation} 
f(\theta,\phi) = f_0\, \exp\left[{- \left(\frac{\phi-\phi_{\rm p}}
                 {\sigma_\phi}\right)^2}\right]~, 
\end{equation} 
where $\phi_{\rm p}$ is the peak location of the Gaussian function and
$f_0$ is the amplitude. If $w_\phi$ is the full width at half-maximum
(FWHM) then $\sigma_\phi= w_\phi/(2\sqrt{\ln 2}). $

Taking into account of modulation, equations (\ref{Is})--(\ref{QUVs}) 
can be written as
\begin{eqnarray}\label{IQUVs}
I_{\rm s} & =& \int^{\theta_{\rm 0}+\delta\theta}_{\theta_{\rm 0}-\delta\theta}
           \int^{\phi_0+\delta\phi}_{\phi_0-\delta\phi} f\,I\, 
             \sin\theta\,d\theta\,d\phi ~\nonumber \\
Q_{\rm s} &=& \int^{\theta_{\rm 0}+\delta\theta}_{\theta_{\rm 0}-\delta\theta}
           \int^{\phi_0+\delta\phi}_{\phi_0-\delta\phi} f\,Q\,
             \sin\theta\,d\theta\,d\phi ~,\nonumber \\
U_{\rm s} &=& \int^{\theta_{\rm 0}+\delta\theta}_{\theta_{\rm 0}-\delta\theta}
           \int^{\phi_0+\delta\phi}_{\phi_0-\delta\phi} f\,U\, 
             \sin\theta\,d\theta\,d\phi ~,\nonumber \\
V_{\rm s} &=& \int^{\theta_{\rm 0}+\delta\theta}_{\theta_{\rm 0}-\delta\theta}
           \int^{\phi_0+\delta\phi}_{\phi_0-\delta\phi} f\,V\, 
             \sin\theta\,d\theta\,d\phi ~.
\end{eqnarray}

\begin{figure}
\epsscale{1.0 }
\plotone{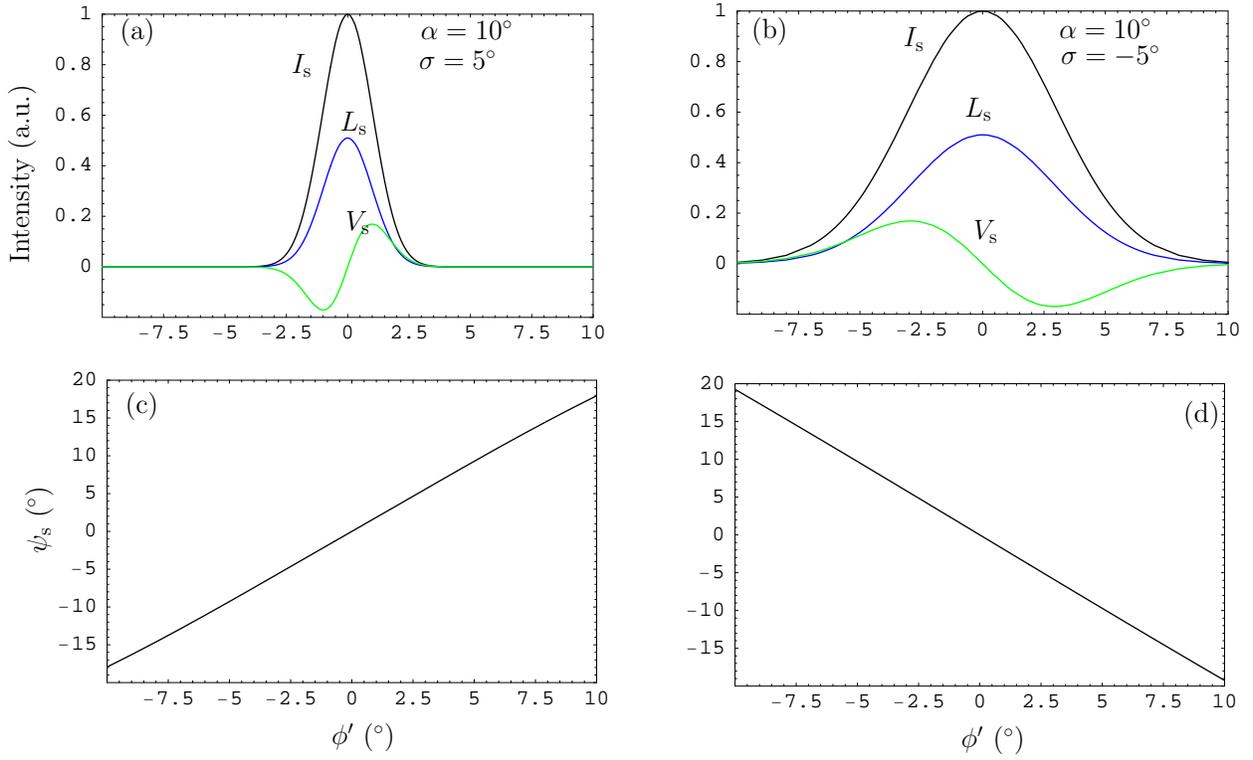}
\caption{Simulated pulse profiles. Given $P=1$~s and $\gamma=400.$
   $\sigma_\phi=0.1,$ $\phi_p=0^\circ,$ and $f_0=1$ are used for the
  modulating Gaussian. }
\label{f6}
\end{figure}

\begin{figure}
\epsscale{1.0 }
\plotone{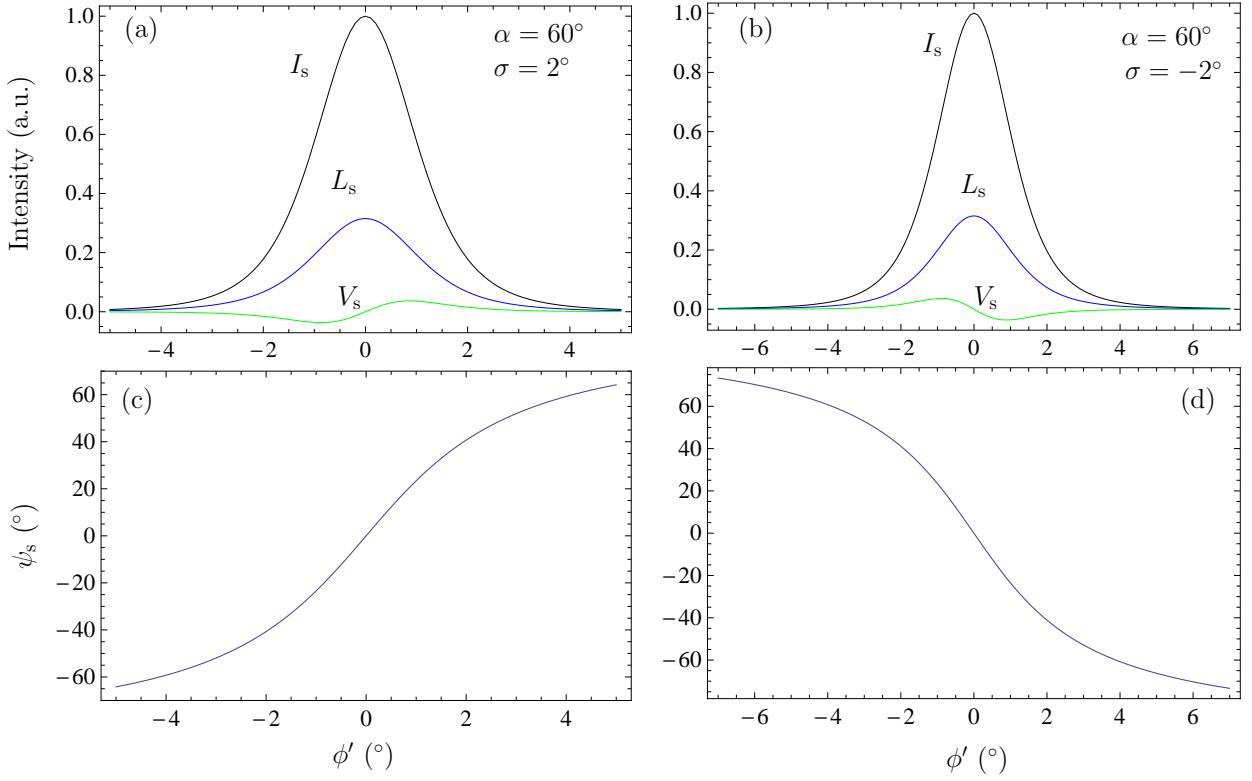} 
\caption{Simulated pulse profiles. Given $P=1$~s and $\gamma=400.$
  For panels (a) and (c) $\sigma_\phi=1,$ $\phi_p=0^\circ, $ and
  $f_0 = 1,$ respectively, are used for the Gaussian. Similarly, for panels
  (b) and (d) $\sigma_\phi=1,$ $\phi_p=180, $ and $f_0 = 1$ are used,
  respectively. }
\label{f7}
\end{figure}

Using a Gaussian with peak located at the meridional plane
$(\phi'=0^\circ),$ we have computed the pulse profiles in the two
cases of impact parameter $(\sigma)$ and inclination angles $(\alpha)$
and plotted in Figures~\ref{f6} and \ref{f7}.  We observe that the
profile in the case of negative $\sigma$ is broader than the positive
case.  This difference is due to the projection of emission region on
to the equatorial plane of the pulsar. In the case of positive
$\sigma,$ the polarization angle $\chi_{\rm s}$ swing is ccw and the
sign change of $V_{\rm s}$ is $-/+$ with respect to $\phi',$ while in
the case of negative $\sigma$ the $\chi_{\rm s}$ swing is cw and the
sign change of $V_{\rm s}$ is $+/-.$ Hence we find that {\it the
  polarization angle swing is correlated with the circular
  polarization sign reversal.}  This correlation is invariant with
respect the stellar spin directions.
 
\begin{figure}
\epsscale{1.0 }
\plotone{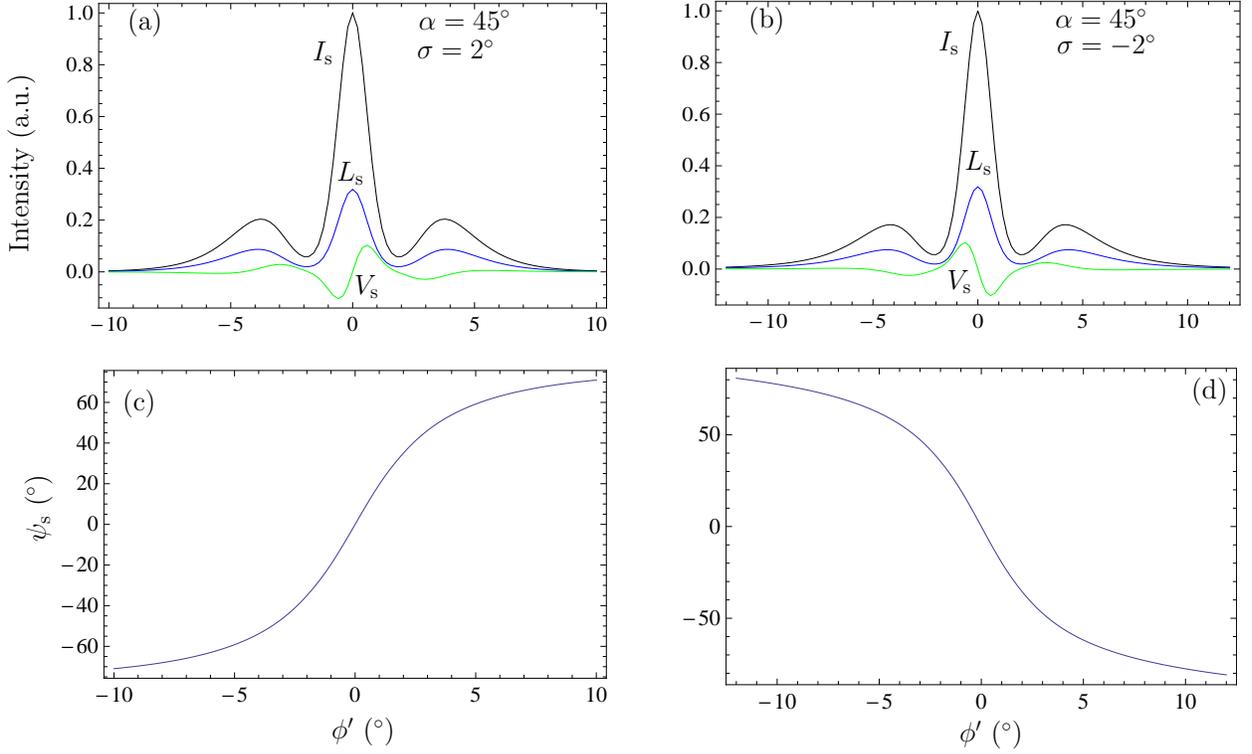} 
\caption{Simulated pulse profiles. Given $P=1$~s and $\gamma=400.$ For
  panels (a) and (c) $\sigma_\phi=0.45,~ 0.32,$ $\phi_p=0^\circ, ~\pm
  60^\circ,$ and $f_0 = 1,~0.9,$ respectively, are used for the
  Gaussians. Similarly, for panels (b) and (d) $\sigma_\phi=0.45,~
  0.32,$ $\phi_p=180^\circ, ~180^\circ\pm 60^\circ,$ and $f_0 =
  1,~0.9,$ respectively, are used from the central component to the
  outer one. }
\label{f8}
\end{figure}

\begin{figure}
\epsscale{1.0 }
\plotone{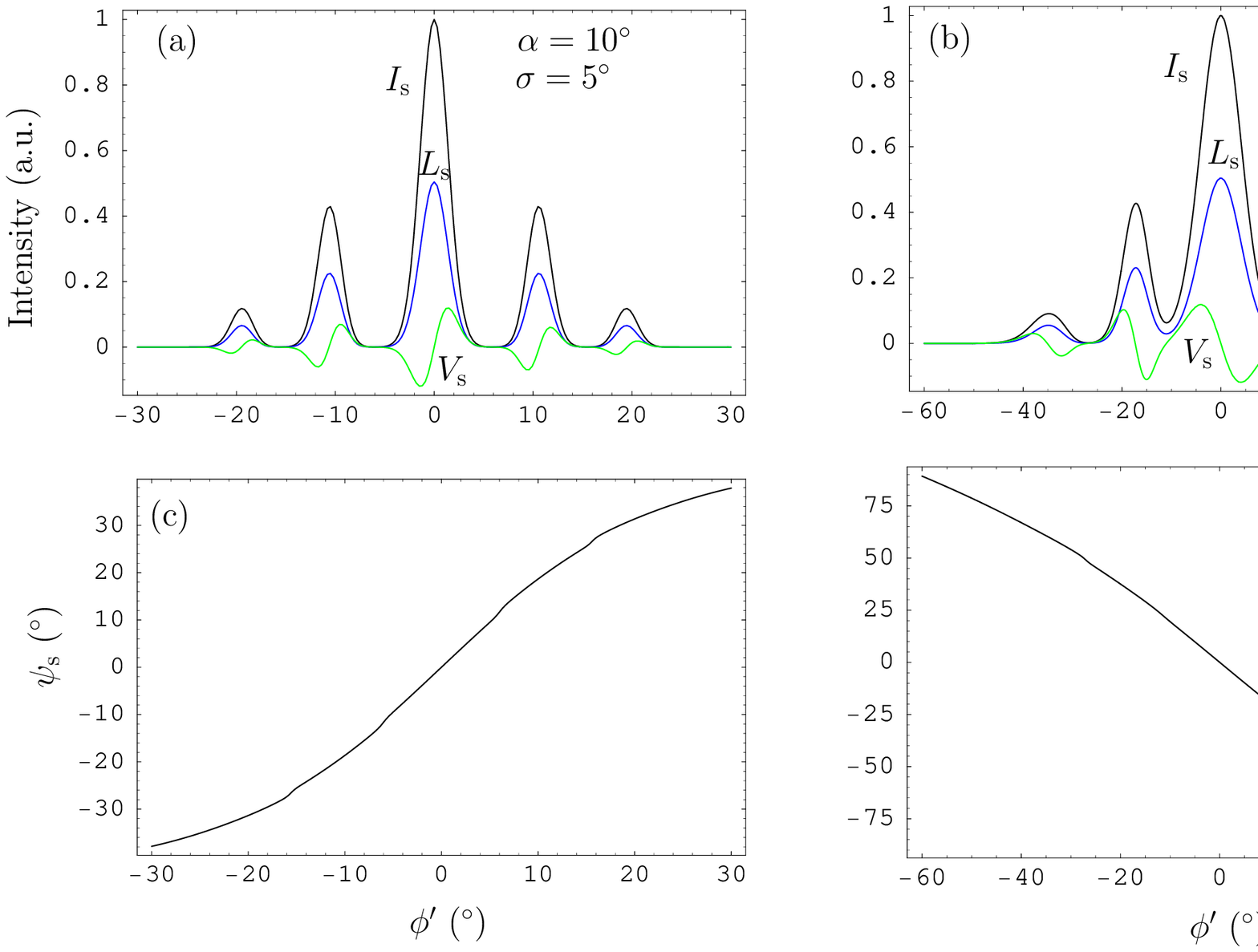}
\caption{Simulated pulse profiles. Given $P=1$~s and $\gamma=400.$ For
  panels (a) and (c) $\sigma_\phi=0.14,~ 0.10, ~0.07,$
  $\phi_p=0^\circ, ~\pm 30^\circ, ~\pm 50^\circ,$ and $f_0 =
  1,~0.75,~0.5,$ respectively, are used for the Gaussians. Similarly,
  for panels (b) and (d) $\sigma_\phi=0.14,~ 0.06, ~0.03,$
  $\phi_p=180^\circ, ~180^\circ\pm 16^\circ, ~180^\circ\pm 26^\circ,$
  and $f_0 = 1,~0.75,~0.5,$ respectively, are used from the central
  component to the outermost one. }
\label{f9}
\end{figure}

\begin{figure}
\epsscale{1.0 }
\plotone{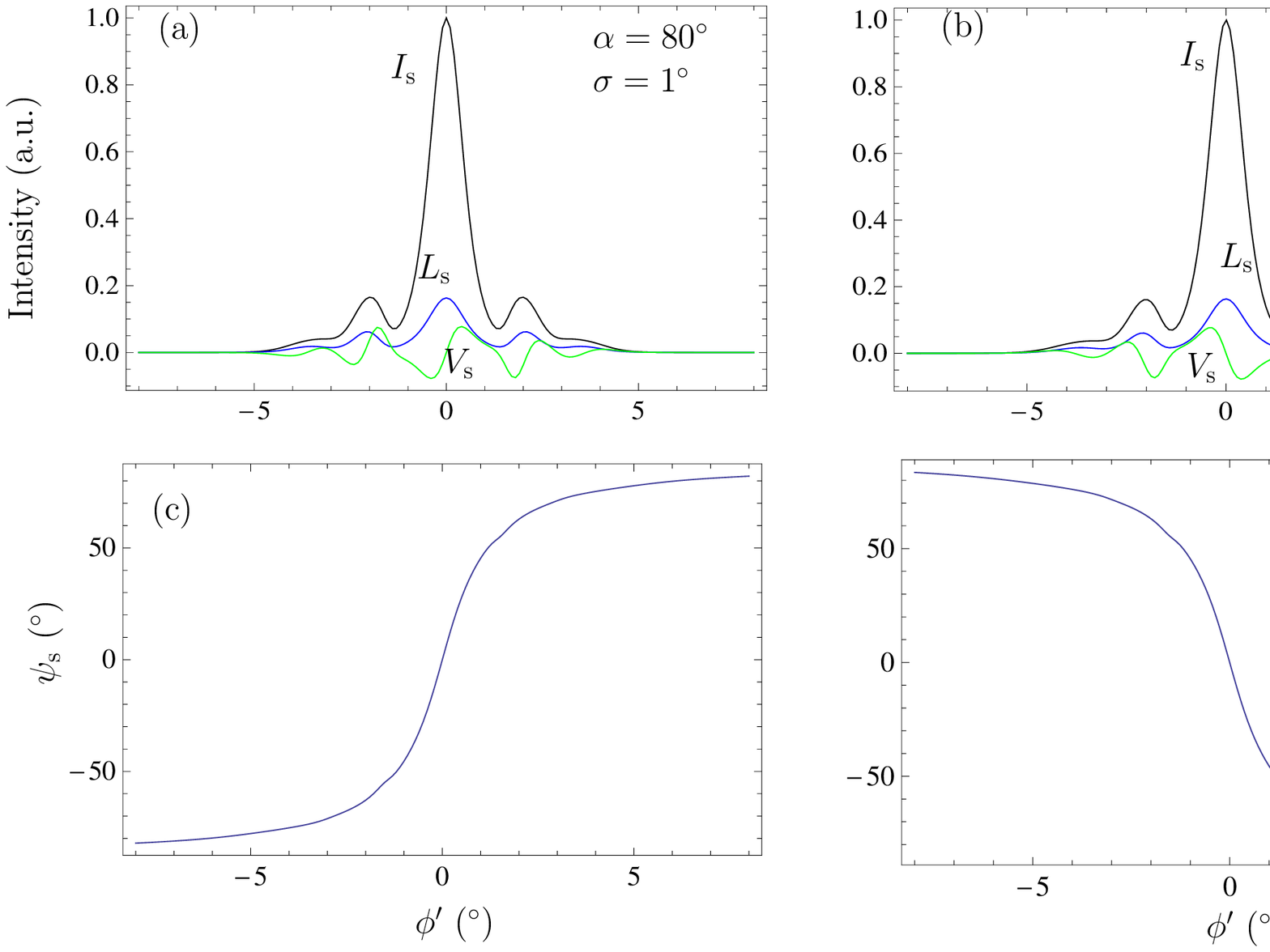}
\caption{Simulated pulse profiles. Given $P=1$~s and $\gamma=400.$ For
  panels (a) and (c) $\sigma_\phi=1,~ 0.1, ~0.05,$ $\phi_p=0^\circ,
  ~\pm 65^\circ, ~\pm 75^\circ,$ and $f_0 = 1,~0.9,~0.8,$
  respectively, are used for the Gaussians. Similarly, for panels (b)
  and (d) $\sigma_\phi=1,~ 0.1, ~0.05,$ $\phi_p=180^\circ,
  ~180^\circ\pm 65^\circ, ~180^\circ\pm 75^\circ,$ and $f_0 =
  1,~0.9,~0.8,$ respectively, are used from the central component to
  the outermost one. }
\label{f10}
\end{figure}

The mean pulsar profiles often found to consist of odd number of
multi-components or subpulses. Many of the works on pulsar profiles
(e.g., Rankin 1990, 1993; Mitra \& Deshpande 1999) have proposed that
the pulsar emission beam has a nested conal structure. To investigate
the polarization of sub-pulses in such profiles, we have reproduced a
five component profile by considering three Gaussians in
Figure~\ref{f8}, and five Gaussians in Figures~\ref{f9} and \ref{f10}.
The central component is presumed to be a core, and the other
components are symmetrically located on either side of the core
forming the cones.  We find across each component that circular
polarization changes the sign and is correlated with the polarization
angle swing. We also observe that the circular polarization of the
outermost components is weaker compared to that of inner ones, which
is quite clear in the case of large inclination angles. This is due to
the fact that the sight line crosses the field lines in the almost
edge-on position in the case of outermost components.  The small
distortions in the polarization angle curve are due to modulation.

\section{Discussion}

Observed pulsar radio luminosities together with the small source size
imply extraordinarily high brightness temperatures, i.e., as high as
$10^{31}$~K. The incoherent sum of a single-particle curvature
radiation is not enough to explain the very high brightness
temperature of pulsar radio emission, therefore, one is forced to
postulate the existence of charged bunches.  To avoid implausibly high
particle densities and energies, coherent radiation processes are
invoked.  Pacini and Rees (1970), and Sturrock (1971) among others
were quick to point out that the observed coherence may be due to
bunching of particles in the emission region of the magnetosphere.
The problem of bunch formation has been known for many decades, and it
has already been addressed by many authors (e.g., Karpman et al. 1975;
Ruderman \& Sutherland 1975; Cox 1979; Asseo, Pelletier \& Sol 1990;
Gil, Lyubarsky \& Melikidze 2004). The natural mechanism for the
formation of charged bunches was first proposed by Karpman et
al. (1975). They have argued that the modulational instability in the
turbulent plasma generates charged solitons, provided that species of
different charges have different masses.  One should mention here that
to explain coherent radio emission we do not necessarily need stable
solitons but only large-scale (as compared with the Langmuir
wavelength) charge density fluctuations.  Gil, Lyubarsky and Melikidze
(2004) generalized the soliton model by including formation and
propagation of the coherent radiation in the magnetospheric plasma.
However, it is not easy to form such charge bunches (see Melrose 1992
for a review). Further, Michel (1991) has pointed out that the
pair-production discharge mechanism originally applied to pulsars by
Sturrock automatically produces dense bunches that can produce
coherence at radio frequencies with sufficient intensity to simulate
pulsar action.  If the bunches of plasma particles with sizes much
smaller than a wavelength of radiation exist then the net radiation
field $\mbox{\boldmath $E$}(\omega)\approx N\mbox{\boldmath $E$}_{\rm
  o}(\omega),$ where $N$ is the number of charges present in the
bunch.  Hence the total radiation field due to a bunch of particles is
equal to the vector sum of the fields radiated by each charge.

In the general framework of models including ours, in which the
radio power is curvature radiation emitted by charge bunches
constrained to follow the field lines, the linear polarization is
intrinsic to the emission mechanism and is, furthermore, a purely
geometric property. In this direction the recently acheieved
observational results and the model predications based on them by
Mitra, Gil and Melikidze (2009) become very relevant. They find that
the polarization angle of linear polarization in subpulses follow
closely the mean polarization angle curve at the corresponding profile
components and argue that their findings favor coherent curvature
radiation over maser mechanism as the observed emission.

 In an actual case, it is the combination of both incoherent and
 coherent superpositions determining the polarization state of the
 observed emission. Though the emissions from a single bunch is highly
 polarized, the radiation received by a distant observer will be less
 polarized, as the radiation from many bunches is incoherently
 superposed (Gil \& Rudnicki 1985).  Also, the degree of polarization
 is found, depending on the time resolution chosen in the observation
 (Gangadhara et~al. 1999).  Circular polarization is generally
 strongest in the central regions of a profile, but is by no means
 confined to these regions. It has been detected from conal components
 of many pulsars, for example, conal-double pulsars and found to be
 highly correlated with the polarization angle swing (You \& Han
 2006).  In most of the cases the sense reversal of circular
 polarization is nearly independent of frequency, suggesting that the
 circular polarization does not arise from propagation or plasma
 effects (Michel 1987; Radhakrishnan \& Rankin 1990).  Radhakrishnan
 and Rankin (1990) have argued that the circular polarization is
 intrinsically antisymmetric type and correlated with the polarization
 angle swing. The antisymmetric circular polarization of curvature
 radiation becomes significant if there are gradients in the
 emissivity over angular scales comparable with the emission cone of
 single charge.  Their results are consistent with the predictions of
 our model (Figures \ref{f6} -- \ref{f10}) and strongly suggest that
 the correlation of antisymmetric circular polarization with the
 polarization angle swing is a geometric property of emission
 mechanism.  Since our model deals with steady flow of relativistic
 plasma bunches along dipolar field lines, it is relevent only for
 average profiles, and reflects the results which are more of
 geometric dependent.  We have not considered any fluctuations or
 instability in the plasma flow. Hence it can not reproduce the
 behaviors of single pulses.

 By adopting the antenna mechanism, Buschauer and Benford (1976) have
 derived a new formalism for the relativistic curvature radiation.
 However, the treatment given does not include the detailed geometry
 of dipolar field lines and the estimation of polarization,
 particularly circular, as we have considered in our model.  Since
 radiation from many bunches is superposed on any given pulse
 longitude, the circular polarization of different signs and
 magnitudes is added. The result of such an addition could be the
 reason for the diversities in the observed circular polarization.
 Since our model was aimed at analyzing the intrinsic polarization
 properties of coherent radiation, we plan to consider the propagation
 effects separately in the subsequent works.  We speculate that the
 propagation origin of antisymmetric circular polarization is very
 unlikely but the symmetric circular polarization may be possible.

\section{Conclusion}  
By taking into account of a detailed geometry of dipolar magnetic
field lines, we have derived the polarization state of the coherent
curvature radiation due to relativistic plasma in the pulsar
magnetosphere, and drawn the following conclusions:
\begin{enumerate}
\item We do confirm the previous results of Gil et~al. (1990a,b; 1993)
  that coherent curvature radiation has basically antisymmetric type
  of circular polarization.  Though the emission from a single bunch
  is highly polarized, the net emission from many bunches within the
  beaming region is less polarized due to the incoherent superposition
  of radiation fields.
 
\item Based on the Stokes parameters of the curvature radiation we
  have deduced the polarization angle swing, i.e., the rotating vector
  model.

\item Based on the coherent curvature radiation, we have achieved for
  the first time the result that {\it the antisymmetric type of
    circular polarization is correlated with the polarization angle
    swing,} and such correlations have been indeed found in the
  profiles of many pulsars.

\item The addition of circular polarization with different signs and
  magnitudes at any given phase could be responsible for the wide
  diversity in circular polarization across the pulse. It is
  consistent with the earlier results (e.g., Gil et~al. 1993, 1995).
\end{enumerate}
 
\acknowledgments

I thank J. L. Han and J. M. Rankin for illuminating discussions.

\renewcommand{\theequation}{A-\arabic{equation}}
\setcounter{equation}{0}  
\section*{APPENDIX~A}
\section*{A.1.~~The Series Expansion Coefficients of Equation~(\ref{coe})}
\begin{eqnarray}   
A_{\rm{x0}} &=& \frac{\cos \zeta\, \left(d_3 N_{\rm x}\left(\theta
  _0\right)+d_2 N_{\rm z}\left(\theta _0\right)\right)}{d_1^2}~,\\
A_{\rm{x1}}&=& \frac{\cos \zeta\, \left(d_1 \left(d_9-f_1\right)-2
  d_3 d_4 N_{\rm x}\left(\theta _0\right)-2 d_2 d_4 N_{\rm z}\left(\theta
  _0\right)\right)}{d_1^3}~,\\
 A_{\rm{x2}}&=& \frac{\left(d_{13} d_1^2-4 d_4 d_{12} d_1-2 d_7
   d_8\right) \cos \zeta\,}{2 d_1^4}~,\\
A_{\rm{x3}}&=& \frac{\left(d_1^3 \left(d_{11}-3
  \left(f_4+f_5\right)-f_6\right)-2 \left(d_6 d_8+3 d_4 d_{13}\right)
  d_1^2+d_{14}\right) \cos \zeta\,}{6 d_1^5}~,\\
A_{\rm{y0}}&=& \frac{d_1 N_{\rm y}\left(\theta _0\right)-e_1 \beta
  _{\rm y}\left(\theta _0\right)}{d_1^2}~,\\
A_{\rm{y1}}&=& \frac{d_1^2 N_{\rm y}'\left(\theta _0\right)-d_1
   e_4 \beta _{\rm y}\left(\theta _0\right)+e_5 N_{\rm y}\left(\theta
   _0\right)+e_1 e_2}{d_1^3} ~,\\
A_{\rm{y2}}&=& \frac{d_1 \left(e_{10}-2 d_1 d_4
  N_{\rm y}'\left(\theta _0\right)\right)+d_1 \left(2 d_4^2-d_1
  d_5\right) N_{\rm y}\left(\theta _0\right)+e_9 \beta _{\rm y}\left(\theta
  _0\right)}{2 d_1^4}~,\\
A_{\rm{y3}}&=& \frac{d_1 \left(3 d_1 \left(d_1 d_5-4 d_4^2-2
  d_7\right) N_{\rm y}'\left(\theta _0\right)+d_1 e_{13}+3
  e_{14}\right)-d_1 e_{11} N_{\rm y}\left(\theta _0\right)+e_{12} \beta
  _{\rm y}\left(\theta _0\right)}{6 d_1^5}~,\nonumber \\ & & \\
A_{\rm{z0}}&=& \frac{\sin \zeta\, \left(-d_3 N_{\rm x}\left(\theta
  _0\right)-d_2 N_{\rm z}\left(\theta _0\right)\right)}{d_1^2}~,\\
A_{\rm{z1}}&=& \frac{\sin \zeta\, \left(d_1 \left(-d_3
  N_{\rm x}'\left(\theta _0\right)-d_2
  N_{\rm z}'\left(\theta _0\right)+f_1\right)+2 d_4 \left(d_3
  N_{\rm x}\left(\theta _0\right)+d_2 N_{\rm z}\left(\theta _0
     \right)\right)\right)}{d_1^3}~,\\
A_{\rm{z2}}&=& \frac{\left(d_1 \left(4 d_4 \left(d_9-f_1\right)+d_1
  \left(-d_{10}+f_2+2 f_3\right)+2 d_5 d_8\right)-6 d_4^2 d_8\right)
  \sin \zeta\,}{2 d_1^4}~,\\
A_{\rm{z3}}&=& \frac{\left(f_7+d_1^3 \left(f_6-d_{11}+3
  \left(f_4+f_5\right)\right)+2 d_1^2 \left(d_6 d_8+3 d_4 d_{10}-3 d_4 \left(f_2+2
  f_3\right)\right)\right) \sin \zeta\,}{6
  d_1^5}~,\nonumber \\ 
& &
\end{eqnarray}
where
\begin{eqnarray} 
 d_1&=& \sin \zeta\, \beta _{\rm x}\left(\theta _0\right)+\cos \zeta\,
        \beta _{\rm z}\left(\theta _0\right)-1~, \\ 
d_2&=& \sin \zeta\,-\beta _{\rm x}\left(\theta _0\right)~, \\ 
d_3&=& \beta _{\rm z}\left(\theta _0\right)-\cos \zeta\,, \\ 
d_4&=& \sin \zeta\, \beta _{\rm x}'\left(\theta _0\right)+
       \cos \zeta\, \beta _{\rm z}'\left(\theta _0\right)~, \\ 
d_5&=& \sin \zeta\, \beta _{\rm x}''\left(\theta _0\right)+
       \cos \zeta\, \beta _{\rm z}''\left(\theta _0\right)~, \\ 
d_6&=& \sin \zeta\, \beta _{\rm x}{}^{(3)}\left(\theta _0\right)+
       \cos \zeta\, \beta _{\rm z}{}^{(3)}\left(\theta _0\right)~, \\ 
d_7&=& d_1 d_5-3 d_4^2~, \\ 
d_8&=& d_3 N_{\rm x}\left(\theta _0\right)+d_2 N_{\rm z}
       \left(\theta _0\right)~, \\ 
d_9&=& d_3 N_{\rm x}'\left(\theta _0\right)+d_2 N_{\rm z}'
       \left(\theta _0\right)~, \\ 
d_{10}&=& d_3 N_{\rm x}''\left(\theta _0\right)+d_2
       N_{\rm z}''\left(\theta _0\right)~, \\ 
d_{11}&=& d_3 N_{\rm x}{}^{(3)}\left(\theta _0\right)+d_2
        N_{\rm z}{}^{(3)}\left(\theta _0\right)~, \\ 
d_{12}&=& d_9-N_{\rm z}\left(\theta _0\right) \beta _{\rm x}'
       \left(\theta _0\right)+N_{\rm x}\left(\theta _0\right) 
       \beta _{\rm z}'\left(\theta _0\right)~, \\ 
d_{13}&=& d_{10} -2 N_{\rm z}'\left(\theta _0\right) \beta _{\rm x}'
      \left(\theta _0\right)+2 N_{\rm x}'\left(\theta _0\right) 
      \beta _{\rm z}'\left(\theta _0\right)-N_{\rm z}\left(\theta _0\right) 
      \beta _{\rm x}''\left(\theta _0\right)+  \\ \nonumber 
   & & N_{\rm x}\left(\theta _0\right)
      \beta _{\rm z}''\left(\theta _0\right)~, \\ 
d_{14}&=& 18 d_1 d_4 d_5 d_8-24 d_8 d_4^3-6 d_1 d_7 d_{12}~,\\
e_1 & = & \sin \zeta\, N_{\rm x}\left(\theta _0\right)+\cos \zeta\,
          N_{\rm z}\left(\theta _0\right)~, \\ 
e_2 & = & 2 d_4 \beta _{\rm y}\left(\theta
         _0\right)-d_1 \beta _{\rm y}'\left(\theta _0\right)~, \\ 
e_3 & = & d_1 \beta _{\rm x}'\left(\theta _0\right)-2 d_4 
          \beta _{\rm x}\left(\theta _0\right)~, \\ 
e_4 & = & \sin \zeta\, N_{\rm x}'\left(\theta _0\right)+\cos \zeta\, 
        N_{\rm z}'\left(\theta _0\right)~, \\ 
e_5 & = & d_1 \cos \zeta\, \beta _{\rm z}'\left(\theta _0\right)-
          2 d_4 \cos \zeta\, \beta _{\rm z}\left(\theta _0\right)+2 d_4+e_3 
        \sin (\zeta )~, \\ 
e_6 & = & \sin \zeta\, N_{\rm x}''\left(\theta _0\right)+\cos \zeta\, 
       N_{\rm z}''\left(\theta _0\right)~, \\ 
e_7 & = & \sin \zeta\, N_{\rm x}{}^{(3)}\left(\theta _0\right)+\cos \zeta\,
       N_{\rm z}{}^{(3)}\left(\theta _0\right)~, \\ 
e_8 & = & d_1 N_{\rm y}{}^{(3)}\left(\theta _0\right)-e_1 \beta _{\rm y}{}^{(3)}
      \left(\theta _0\right)~, \\ 
e_9 & = & 4d_1 d_4 e_4-6 d_4^2 e_1+d_1 \left(2 d_5 e_1-d_1 e_6\right)~, \\ 
e_{10} & = & d_1 \left(d_1 N_{\rm y}'' \left(\theta _0\right)-e_1 \beta _{\rm y}''
            \left(\theta _0\right)\right)+\left(4 d_4 e_1-2 d_1 e_4\right) 
            \beta _{\rm y}'\left(\theta _0\right)~, \\ 
e_{11} & = & 24 d_4^3+6d_4 \left(d_7-2 d_1 d_5\right) +d_1^2 d_6~, \\ 
e_{12} & = & 24 d_4^3 e_1+6 d_1 d_4 \left(d_1 e_6-3 d_5 e_1\right)+d_1 \left(-d_1^2
             e_7+2 d_6 d_1 e_1+6 d_7 e_4\right)~, \\ 
e_{13} & = & \left(6 d_4 e_1-3 d_1 e_4\right) \beta _{\rm y}''\left(\theta _0\right)-3 
             d_1 d_4 N_{\rm y}''\left(\theta _0\right)+d_1 e_8~, \\ 
e_{14} & = & \left(2 d_7 e_1+d_1 \left(4 d_4 e_4-d_1 e_6\right)\right) \beta _{\rm y}'
             \left(\theta _0\right)~, \\
f_1 & = & N_{\rm z}\left(\theta _0\right) \beta _{\rm x}'\left(\theta _0\right)-N_{\rm x}
          \left(\theta _0\right) \beta _{\rm z}'\left(\theta _0\right)~, \\ 
f_2 & = & N_{\rm z}\left(\theta _0\right) \beta _{\rm x}''\left(\theta _0\right)-
          N_{\rm x}\left(\theta _0\right) \beta _{\rm z}''\left(\theta _0\right)~, \\ 
f_3 & = & N_{\rm z}'\left(\theta _0\right) \beta _{\rm x}'\left(\theta _0\right)-N_{\rm x}'
          \left(\theta _0\right) \beta _{\rm z}'\left(\theta _0\right)~, \\ 
f_4 & = & N_{\rm z}'\left(\theta _0\right) \beta _{\rm x}''\left(\theta _0\right)-N_{\rm x}'
          \left(\theta _0\right) \beta _{\rm z}''\left(\theta _0\right)~, \\ 
f_5 & = & N_{\rm z}''\left(\theta _0\right) \beta _{\rm x}'\left(\theta _0\right)-N_{\rm x}''
          \left(\theta _0\right) \beta _{\rm z}'\left(\theta _0\right)~, \\ 
f_6 & = & N_{\rm z}\left(\theta _0\right) \beta _{\rm x}{}^{(3)}\left(\theta _0\right)-
          N_{\rm x}\left(\theta _0\right) \beta _{\rm z}{}^{(3)}\left(\theta _0\right)~, \\ 
f_7 & = & 24 d_4^3 d_8-6 d_1 \left(d_7 \left(f_1-d_9\right)+3 d_4 d_5 d_8\right) ~.
\end{eqnarray}
 The expression of \mbox{\boldmath $\beta$}$=\{\beta_{\rm
   x},\,\beta_{\rm y} ,\,\beta_{\rm z}\}$ and the derivatives
 $\mbox{\boldmath $\beta$} ',\,\mbox{\boldmath $\beta$}'',\,
 \mbox{\boldmath $\beta$}^{(3)}$ and \mbox{\boldmath $\beta$}$^{(4)}$,
 which respectively represent the first, second, third, and fourth
 order differentiations with respect to $\theta$ evaluated at
 $\theta_0,$ are as follows:
\begin{eqnarray}
\mbox{\boldmath $\beta$} (\theta _0)\!\!\! &=&\!\!\! \Big\{\kappa (h_1 \sin (\Gamma
   (\theta _0))+h_3 \cos (\Gamma (\theta _0))),~
   \kappa (h_2 \sin (\Gamma (\theta _0))+h_4 
   \cos (\Gamma (\theta _0))),\nonumber\\ 
   &&\kappa (\cos \alpha\,  
   \cos (\Gamma (\theta _0))-h_5 \sin (\Gamma
   (\theta _0)))\Big\} ~,\label{A48} \\
\mbox{\boldmath $\beta$} '(\theta _0)\!\!\!&=&\!\!\! \Big\{\kappa  \Gamma '(\theta _0) 
   (h_1 \cos (\Gamma (\theta _0))-h_6
   \sin (\Gamma (\theta _0))),~\kappa  \Gamma '
   (\theta _0) (h_2 \cos (\Gamma (\theta _0))-\nonumber\\ &&
   h_4 \sin (\Gamma (\theta _0))),~\kappa  \Gamma '
   (\theta _0) (-h_5 \cos (\Gamma (\theta _0))-
   \cos \alpha\,  \sin (\Gamma (\theta _0)))\Big\}~, \\    
\mbox{\boldmath $\beta$} ''(\theta _0)\!\!\!&=&\!\!\! \Big\{\kappa  (\Gamma '(\theta _0){}^2 
   (h_8 \sin (\Gamma (\theta _0))-\cos \phi '\, 
   (h_9 \sin (\Gamma (\theta _0))+\sin \alpha\,  
   \cos (\Gamma (\theta _0))))+ \nonumber\\ 
&& \Gamma ''(\theta _0) (h_1 \cos (\Gamma (\theta _0))-
   h_6 \sin (\Gamma (\theta _0)))),~
 \kappa (\Gamma ''(\theta _0) (h_2 \cos (\Gamma
   (\theta _0))-\nonumber\\ 
&& h_4 \sin (\Gamma (\theta _0)))
   +\Gamma '(\theta _0){}^2 (h_2 (-\sin (\Gamma 
   (\theta _0)))-h_4 \cos (\Gamma (\theta _0)) ))~,\nonumber\\ 
&& \kappa (\Gamma '(\theta _0){}^2 (h_5 \sin (\Gamma
   (\theta _0))-\cos \alpha\,  \cos (\Gamma
   (\theta _0)))-\Gamma ''(\theta _0)
   (h_5 \cos (\Gamma (\theta _0))+\nonumber\\ 
&&\cos \alpha\,   \sin (\Gamma (\theta _0))))\Big\} ~,   \\ 
\mbox{\boldmath $\beta$} ^{(3)}(\theta _0)\!\!\!&=&\!\!\! \Big\{\kappa  (\Gamma ^{(3)}
   (\theta _0) (h_1 \cos (\Gamma (\theta _0))-
   h_6 \sin (\Gamma (\theta _0)))+
\Gamma '(\theta _0){}^3 (h_6 \sin (\Gamma (\theta _0)- \nonumber\\
&& h_9\cos (\Gamma (\theta _0)) \cos \phi ')+h_8 \cos
   (\Gamma (\theta _0)))-3 \Gamma '(\theta _0) 
   \Gamma ''(\theta _0) (h_1 \sin
   (\Gamma (\theta _0))+\nonumber\\ 
&& h_3 \cos (\Gamma (\theta _0)))),~\kappa  (\Gamma ^{(3)}
   (\theta _0) (h_2 \cos (\Gamma (\theta _0))-
   h_4 \sin (\Gamma (\theta _0)))+\nonumber\\ 
&& \Gamma '(\theta _0) {}^3 (h_4
   \sin (\Gamma (\theta _0))-h_2 \cos (\Gamma
   (\theta _0)))-3 \Gamma '(\theta _0)
   \Gamma ''(\theta _0) (h_2 \sin (\Gamma
   (\theta _0))+\nonumber\\ 
&&h_4 \cos (\Gamma 
   (\theta _0)))),~\kappa  (-\Gamma ^{(3)}
   (\theta _0) (h_5 \cos (\Gamma (\theta _0))+
   \cos \alpha\,  \sin (\Gamma (\theta _0)))+\nonumber\\ 
&& \Gamma '(\theta _0){}^3 (h_5
   \cos (\Gamma (\theta _0))+\cos \alpha\,  \sin
   (\Gamma (\theta _0)))+3 \Gamma '(\theta _0) 
   \Gamma ''(\theta _0) (h_5 \sin
   (\Gamma (\theta _0))-\nonumber\\  
&& \cos \alpha\,  \cos (\Gamma (\theta _0))))\Big\}~,\\   
\mbox{\boldmath $\beta$} ^{(4)}(\theta _0)\!\!\! &=&\!\!\! \Big\{\kappa  (h_9 \cos
   \phi '\, (\sin (\Gamma (\theta _0)) 
   (-3 \Gamma ''(\theta _0){}^2+\Gamma '(\theta _0){}^4-4 
   \Gamma ^{(3)}(\theta _0)
   \Gamma '(\theta _0))+\nonumber\\  
&& \cos (\Gamma (\theta _0)) 
   (\Gamma ^{(4)}(\theta _0)-6 \Gamma '(\theta _0){}^2 
   \Gamma ''(\theta _0)))+h_8 (\sin (\Gamma 
   (\theta _0)) (3 \Gamma ''(\theta _0){}^2-\nonumber\\  
&&  \Gamma '(\theta _0){}^4+4 \Gamma ^{(3)}(\theta _0)
   \Gamma '(\theta _0))+\cos (\Gamma (\theta _0)) 
   (6 \Gamma '(\theta _0){}^2 \Gamma ''(\theta _0)-
   \Gamma ^{(4)}(\theta _0)))+\nonumber\\  
&& h_6 (\sin (\Gamma 
   (\theta _0)) (6 \Gamma '(\theta _0){}^2 
   \Gamma ''(\theta _0)-\Gamma ^{(4)}(\theta _0))+
   \cos (\Gamma (\theta _0))
   (-3 \Gamma ''(\theta _0){}^2+\nonumber\\  
&&\Gamma '(\theta _0){}^4-4 
   \Gamma ^{(3)}(\theta _0) \Gamma '(\theta _0)))),~
   \kappa  (h_2 \sin (\Gamma (\theta _0)) 
   (-3 \Gamma ''(\theta _0){}^2+\Gamma '(\theta _0){}^4-\nonumber\\  
&& 4\Gamma ^{(3)}(\theta _0) \Gamma '(\theta _0))+h_2 
   \cos (\Gamma (\theta _0))
   (\Gamma ^{(4)}(\theta _0)-6 \Gamma '(\theta _0){}^2 
   \Gamma ''(\theta _0))+\nonumber\\  
&& h_4 (\sin
   (\Gamma (\theta _0)) (6 \Gamma '(\theta _0){}^2 
   \Gamma ''(\theta _0)-\Gamma ^{(4)}(\theta _0))+ 
   \cos (\Gamma (\theta _0))
   (-3 \Gamma ''(\theta _0){}^2+\nonumber\\  
&& \Gamma '(\theta _0){}^4-4 
   \Gamma ^{(3)}(\theta _0) \Gamma '(\theta _0)))),~
   \kappa  (\cos \alpha\,  (\sin (\Gamma (\theta _0)) 
   (6 \Gamma '(\theta _0){}^2 \Gamma ''(\theta _0)-\nonumber\\  
&&   \Gamma ^{(4)}(\theta _0))+
   \cos (\Gamma (\theta _0)) 
   (-3 \Gamma ''(\theta _0){}^2+\Gamma '(\theta _0){}^4-4 
   \Gamma ^{(3)}(\theta _0)\Gamma '(\theta _0)))-\nonumber\\  
&&   h_5 (\sin (\Gamma (\theta _0)) 
   (-3 \Gamma ''(\theta _0){}^2+\Gamma '(\theta _0){}^4-4
   \Gamma ^{(3)}(\theta _0) \Gamma '(\theta _0))+\nonumber\\  
&&   \cos (\Gamma (\theta _0))
   (\Gamma ^{(4)}(\theta _0)-6 \Gamma '(\theta _0){}^2 
   \Gamma ''(\theta _0))))\Big\}~,\label{A52} 
\end{eqnarray} 
 where
\begin{eqnarray}  
h_1 & = & \cos \alpha\,  \cos \phi\, \cos \phi '\,-\sin \phi\, 
          \sin \phi '~, \\ 
h_2 & = & \cos \alpha\,  \cos \phi\, \sin \phi '\,+\sin \phi\, 
          \cos \phi '\,, \\ 
h_3 & = & \sin \alpha\,  \cos \phi '~, \\ 
h_4 & = & \sin \alpha\,  \sin \phi '~, \\ 
h_5 & = & \sin \alpha\,  \cos \phi~, \\ 
h_6 & = & \sin \alpha\,  \cos \phi '~, \\ 
h_7 & = & \sin \alpha\,  \sin \phi '~, \\ 
h_8 & = & \sin \phi\, \sin \phi '~, \\ 
h_9 & = & \cos \alpha\,  \cos \phi~, \\
\cos (\Gamma (\theta _0)) &=& \frac{1+3 \cos \left(2 \theta _0\right)}
     {\sqrt{10+6 \cos \left(2 \theta _0\right)}}~,\\
\sin (\Gamma (\theta _0))&=& \frac{3 \sin \left(2 \theta _0\right)}
     {\sqrt{10+6 \cos \left(2 \theta _0\right)}}~, \\ 
\Gamma \left(\theta _0\right)&=& \tan ^{-1}\left(\frac{3 \sin 
   \left(2 \theta _0\right)}{1+3 \cos \left(2 \theta _0\right)}\right)~,\\
\Gamma '\left(\theta _0\right)&=& \frac{4}{5+3 \cos \left(2 \theta _0\right)}+1 ~,\\
\Gamma ''\left(\theta _0\right)&=& \frac{24 \sin \left(2 \theta _0\right)}
    {\left(5+3 \cos \left(2 \theta _0\right)\right){}^2}~,\\
\Gamma ^{(3)}\left(\theta _0\right)&=& \frac{24 \left(10 \cos 
   \left(2 \theta _0\right)-3 \cos \left(4 \theta _0\right)+9\right)}
   {\left(5+3 \cos \left(2 \theta _0\right)\right){}^3}~,\\
 \Gamma ^{(4)}\left(\theta _0\right)&=& \frac{24 \left(107 \sin
   \left(2 \theta _0\right)+120 \sin \left(4 \theta _0\right)-9 \sin
   \left(6 \theta _0\right)\right)}{\left(5+3 \cos \left(2 \theta _0\right)\right){}^4}~.
\end{eqnarray} 
 Note that the $\phi'$, which appears in the above equations, is just
 a variable for the rotation phase, and the prime $(')$ on it does not
 represent any differentiation. 
 
By having known the derivatives of \mbox{\boldmath $\beta$} from
equations~(\ref{A48})--(\ref{A52}), we can define the derivatives
\mbox{\boldmath $N$} evaluated at $\theta_0:$
\begin{eqnarray} 
\mbox{\boldmath $N$}(\theta_0)&=&\mbox{\boldmath $\beta$}'(\theta_0)~,\\
\mbox{\boldmath $N$}'(\theta_0)&=&\mbox{\boldmath $\beta$}''(\theta_0)~,\\
\mbox{\boldmath $N$}''(\theta_0)&=&\mbox{\boldmath $\beta$}^{(3)}(\theta_0)~,\\
\mbox{\boldmath $N$}^{(3)}(\theta_0)&=&\mbox{\boldmath $\beta$}^{(4)}(\theta_0)~.
\end{eqnarray} 
\section*{A.2.~~The series expansion coefficients of equation~(\ref{exp})}
\begin{eqnarray} 
 c_0\!\!\! &=&\!\!\! g_1 \left(2 g_3 \sin ^3\left(\theta _0\right)-3 g_5 \cos
   \left(\theta _0\right)+2 g_2+\sqrt{3} \left(\log (2)-2 \log
   \left(g_4\right)\right)\right)~,      \\
c_1\!\!\! &=&\!\!\! 3 g_1 \Big(\sin \left(\theta _0\right) \left(g_3 \sin \left(2
   \theta _0\right)+2 \sqrt{10+6 \cos \left(2 \theta _0\right)}\right)+ \nonumber \\
&& \kappa  \cos \left(\Gamma _0\right) \left(\sin
   \left(\theta _0\right)-3 \sin \left(3 \theta _0\right)\right)\Big)~,     \\
c_2\!\!\! &=&\!\!\! \frac{3 g_1 \left(g_6 \left(1+3 \cos \left(2 \theta _0\right)\right)+
    2 \kappa  \cos \left(\Gamma _0\right) \left(5+3 \cos
   \left(2 \theta _0\right)\right) \left(\cos \left(\theta _0\right)-9
   \cos \left(3 \theta _0\right)\right)\right)}
    {4 \left(5+3 \cos \left(2 \theta _0\right)\right)} ~,\nonumber \\ 
&&   \\
c_3\!\!\! &=&\!\!\! \frac{1}{4} g_1 \left(-g_3 \cos \left(\theta _0\right)+g_7-\frac{4
   \sqrt{2} \left(28 \sin \left(\theta _0\right)+9 \left(5 \sin \left(3
   \theta _0\right)+\sin \left(5 \theta _0\right)\right)\right)}{\left(5+3
   \cos \left(2 \theta _0\right)\right){}^{3/2}}\right)   ~,
\end{eqnarray} 
where
\begin{eqnarray}     
\cos \left(\Gamma _0\right) & = &  \sin \alpha\,  \sin \zeta\, \cos
           \phi '\,+\cos \alpha\,  \cos \zeta~, \\ 
g_1 & = &  \frac{\omega\, r_e}{12 c \kappa }~, \\ 
g_2 & = &  6+\sqrt{3} \log \left(2+\sqrt{3}\right)~, \\ 
g_3 & = &  6 \kappa  \left(\sin \zeta\, \left(\sin \phi\, \sin 
          \phi '\,-\cos \alpha\,  \cos \phi\, \cos
          \phi '\,\right)+\sin \alpha\,  \cos \zeta\,    
          \cos \phi\,\right)~, \\ 
g_4 & = & \sqrt{6} \cos \left(\theta _0\right)+\sqrt{5+3 \cos
          \left(2 \theta _0\right)}~, \\ 
g_5 & = &  4 \kappa  \cos \left(\Gamma _0\right) \sin ^2\left(\theta _0\right)+  
         \sqrt{10+6 \cos \left(2 \theta _0\right)}~, \\ 
g_6 & = &  7 g_3 \sin \left(\theta _0\right)+3 g_3 \sin \left(3 \theta _0\right)+
           8 \sqrt{10+6 \cos \left(2 \theta _0\right)} \cos \left(\theta _0\right)~, \\ 
g_7 & = &  9 g_3 \cos \left(3 \theta _0\right)-2 \kappa  \cos \left(\Gamma _0\right) 
          \left(\sin \left(\theta _0\right)-27 \sin \left(3 \theta _0\right)\right)~.  
\end{eqnarray} 
\newpage
\renewcommand{\theequation}{B-\arabic{equation}}
\setcounter{equation}{0}  
\section*{APPENDIX~B}
\section*{B.1.~~To Find Solution to Integrals in Equation~(\ref{Ecomp})}

Consider the integral
\begin{equation}
S_0 = \!\int\limits_{-\infty}^{+\infty} {e}^{i(c_{\rm 1} \,\mu+c_{\rm 2} 
      \,\mu^2+c_{\rm 3}\,\mu^3} d\mu~.
\end{equation}
By changing the variable of integration $\mu=(x/l)+m,$ and defining
the constants $l = \sqrt[3]{c_3}$ and $m = -c_2/(3c_3),$ we obtain
\begin{equation}\label{I0}
\!\int\limits_{-\infty}^{+\infty} {e}^{i(c_{\rm 1} \,\mu+c_{\rm 2} 
        \,\mu^2+c_{\rm 3} \,\mu^3)}d\mu 
      = U \!\int\limits_{-\infty}^{+\infty} e^{i \left(z\, x +x^3\right)}dx~, 
\end{equation}
where
$$z = \frac{1}{\sqrt[3]{c_3}}\left(c_1-\frac{c_2^2}{3 c_3}\right), \quad
U = \frac{1}{\sqrt[3]{c_3}} e^{i \frac{c_2}{3 c_3}\left(\frac{2 c_2^2}
    {9 c_3}-c_1\right)}. $$
For ${\rm Im}(z) = 0$ we know
\begin{equation}
j_0 = \int\limits_{-\infty}^{\infty}{e}^{i(z x+x^3)} dx 
    = \frac{\pi}{\sqrt[3]{3}} \left[\left(1-\frac{\sqrt{z^2}}{z}\right)
      \mbox{Ai}\left(-\frac{\sqrt{z^2}}{\sqrt[3]{3}}\right)+\left(1+
      \frac{\sqrt{z^2}}{z}\right)
      \mbox{Ai}\left(\frac{\sqrt{z^2}}{\sqrt[3]{3}}\right)\right]~,
\end{equation}
where Ai$(z)$ is an entire Airy function of $z$ with no branch cut 
discontinuities, and
\begin{equation}
j_1 = \int\limits_{-\infty}^{\infty} x\,{e}^{i(z x+x^3)}dx
    = -i\frac{2 \pi}{\sqrt[3]{3^2}} \mbox{Ai}'\left(\frac{z}{\sqrt[3]{3}}\right)~,
\end{equation}
where Ai$'(z)$ is the derivative of the Airy function Ai$(z).$
Therefore, we have
\begin{equation}\label{io}
S_0 = U\, j_0~.
\end{equation}
By differentiating equation~(\ref{I0}) on both sides with respect to
$c_1$ we obtain
\begin{eqnarray}
S_1 = \!\int\limits_{-\infty}^{+\infty}\mu {e}^{i(c_{\rm 1} \,\mu+c_{\rm 2} 
\,\mu^2+c_{\rm 3} \,\mu^3)}d\mu
  & = & \frac{U}{\sqrt[3]{c_3}}\!\int\limits_{-\infty}^{+\infty}
        \left(x-\frac{c_2}{3\sqrt[3]{c_3^2}}\right) 
        e^{i \left(z\, x +x^3\right)}dx~\nonumber \\
  & = &\frac{U}{\sqrt[3]{c_3}}\left(j_1-\frac{c_2}{3\sqrt[3]{c_3^2}}j_0\right) ~.
\end{eqnarray}
Differentiation of equation~(\ref{I0}) on both sides with respect to
$c_2$ gives
\begin{eqnarray}
S_2 = \int\limits_{-\infty}^{+\infty}\mu^2 {e}^{i(c_{\rm 1} 
        \,\mu+c_{\rm 2} \,\mu^2+c_{\rm 3} \,\mu^3)}d\mu
  & = & \frac{U}{3c_3} \!\int\limits_{-\infty}^{+\infty}\left(\frac{2 c_2^2}{3c_3}-
        c_1-\frac{2c_2}{\sqrt[3]{c_3}}x\right) 
        e^{i \left(z\, x +x^3\right)}dx~\nonumber \\
  & = & \frac{U}{3c_3}\left[\left(\frac{2 c_2^2}{3c_3}-c_1\right)j_0-
        \frac{2c_2}{\sqrt[3]{c_3}}j_1\right]~. 
\end{eqnarray}
Next, by differentiating equation~(\ref{I0}) on both sides with
respect to $c_3$ we obtain
\begin{eqnarray}
S_3 = \!\int\limits_{-\infty}^{+\infty}\mu^3 {e}^{i(c_{\rm 1} 
      \,\mu+c_{\rm 2} \,\mu^2+c_{\rm 3} \,\mu^3)}d\mu
  & = & \frac{U}{9\sqrt[3]{c_3^7}}\!\int\limits_{-\infty}^{+\infty}
      \left(\frac{9 c_1 c_2 c_3-4 c_2^3+i 9 c_3^2}{3 \sqrt[3]{c_3^2}}+
      (4 c_2^2-3 c_1 c_3)x \right) e^{i \left(z\, x +x^3\right)}dx~\nonumber \\
  & = & \frac{U}{9\sqrt[3]{c_3^7}}\left[\frac{(9 c_1 c_2 c_3-4 c_2^3+i 9 c_3^2)}
      {3 \sqrt[3]{c_3^2}}j_0+ (4 c_2^2-3 c_1 c_3)j_1 \right] ~.
\end{eqnarray}



\end{document}